\documentclass{jpp}
\usepackage{graphicx}
\graphicspath{{figures/}}

\usepackage[utf8]{inputenc}
\usepackage[T1]{fontenc}
\usepackage{amsmath}
\usepackage{xcolor}
\usepackage{wrapfig}
\usepackage{soul}

\newcommand{\xt}{\tilde{x}}
\newcommand{\zt}{\tilde{z}}

\shorttitle{tearing instability with jet}
\shortauthor{C. Shi}

\title{Instabilities in a current sheet with plasma jet}

\author{Chen Shi\aff{1}
  \corresp{\email{cshi1993@ucla.edu}}}

\affiliation{\aff{1}Department of Earth, Planetary, and Space Sciences, University of California, Los Angeles, \\ Los Angeles,
CA 90095, USA}

\begin{document}

\maketitle

\begin{abstract}
We study the stability problem of a magnetohydrodynamic current sheet with the presence of a plasma jet. The flow direction is perpendicular to the normal of the current sheet and we analyze two cases: (1) The flow is along the anti-parallel component of the magnetic field; (2) The flow is perpendicular to the anti-parallel component of the magnetic field. A generalized equation set with the condition of incompressibility is derived and solved as a boundary-value-problem. For the first case, we show that, the streaming kink mode is stabilized by the magnetic field at $V_0/B_0 \lesssim 2$, where $V_0$ and $B_0$ are the jet speed and upstream Alfv\'en speed, and it is not affected by resistivity significantly. The streaming sausage mode is stabilized at $V_0/B_0 \lesssim 1$, and it can transit to the streaming tearing mode with a finite resistivity. The streaming tearing mode has larger growth rate than the pure tearing mode, though the scaling relation between the maximum growth rate and the Lundquist number remains unchanged. When the jet is perpendicular to the anti-parallel component of the magnetic field, the most unstable sausage mode is usually perpendicular (wave vector along the jet) without a guide field. But with a finite guide field, the most unstable sausage mode can be oblique, depending on the jet speed and guide field strength. 
\end{abstract}

\section{Introduction}
For nearly one century, scientists have been investigating the mechanisms that cause space weather events such as the magnetic storms \citep[e.g.][]{chapman1940theory,ferraro1952theory}. Magnetic reconnection is identified as one of the most important processes in space plasmas that drives various explosive phenomena, such as the solar flares \citep[e.g.][]{masuda1994loop}, coronal mass ejections \citep[e.g.][]{gosling1995three}, and the magnetic substorms in the Earth's magnetotail \citep[e.g.][]{angelopoulos2008tail}. It efficiently converts the magnetic energy in a current sheet to the kinetic and internal energies of the plasma through re-configuration of the topology of the magnetic field and connectivity of the magnetic field lines. In the laboratory, magnetic reconnection destabilizes the plasma \citep[e.g.][]{yamada1994investigation,dorfman2013three} and is fatal to stable controlled fusion. Understanding of how the reconnection triggers and evolves is crucial for a complete description of how energy is stored and released in different plasma environments.

As the plasma is a multi-scale system, various types of waves and instabilities exist on the largest magnetohydrodynamic (MHD) scale, the intermediate ion-kinetic scale, and the smallest electron-kinetic scale. Consequently, magnetic reconnection is a multi-scale process. In the past two decades, numerical simulations \citep[e.g.][]{daughton2006fully,guo2015particle,cassak2017effect,lu2019turbulence} as well as in-situ measurements of the space plasma \citep[e.g.][]{burch2016electron,torbert2018electron} have greatly enhanced our understanding of the microscopic kinetic physics of reconnection in the ion diffusion region and electron diffusion region. But it still remains unclear as how fast reconnection triggers in the macroscopic current sheets whose dimensions are much larger than any ion kinetic scales, e.g. current sheets in the pre-flare configurations in the solar corona.

In ideal-MHD regime, the magnetic field lines are ``frozen in'' the plasma and change of their connectivity is prohibited. Thus, a certain mechanism that breaks the ideal-MHD condition must play a role for the reconnection to happen. In macroscopic current sheets, this mechanism is either collision-induced resistivity, or some kind of effective resistivity caused by microscopic wave-particle interactions \citep[e.g.][]{buchner2006anomalous,ma2018effective}. Hence, the triggering problem of reconnection at MHD scales is essentially the stability problem of the resistive current sheet. Since 1960s, many works have been conducted on the resistive instability, i.e. the so-called ``tearing instability'', of the current sheet \citep{furth1963finite,coppi1966resistive}. The tearing mode grows with the help of resistivity that transfers the magnetic energy stored in the shear magnetic field to the growing perturbations, leading to the formation of a chain of plasmoids. Considering an infinitely long current sheet with thickness $a$, one can define the dimensionless \textit{Lundquist number} $S= aV_A/\eta$, where $V_A = B/\sqrt{\mu_0 \rho}$ is the characteristic Alfv\'en speed with $\rho$ and $B$ being the plasma density and asymptotic magnetic field strength, and $\eta$ is the magnetic diffusivity, which is resistivity divided by the permeability $\mu_0$. For simplicity, we will refer to $\eta$ as ``resistivity'' hereinafter. Linear theory predicts that the most unstable tearing mode has a growth rate $\gamma \tau_a \sim S^{-1/2}$ where $\tau_a = a/V_A$ is the Alfv\'en crossing time. This relation implies a faster growth of the instability with larger resistivity. As plasma in most of the space environments and laboratories is weakly collisional \citep{ji2011phase,pucci2017fast}, the tearing mode seems to grow at very slow speed. 

However, considering a two-dimensional current sheet whose length is $L$, in most pre-reconnection configurations, its aspect ratio can be very large ($L \gg a$). In this case, we should use $L$ instead of $a$ to measure the growth rate of tearing instability. After re-defining the Alfv\'en crossing time and Lundquist number such that $\tau_L = L/V_A$ and $S_L = LV_A/\eta$, it can be shown that the maximum growth rate of tearing mode is $\gamma \tau_L \sim S_L^{-1/2} \times (a/L)^{-3/2}$ \citep{pucci2013reconnection}, implying that the aspect ratio of the current sheet is an important factor in determining how fast the mode grows. The growth rate can be extremely large at low-resistivity limit ($S_L \rightarrow \infty$) if the current sheet is thinner than a critical value $a/L \sim S_L^{1/3}$. Especially, in the classic model for steady reconnection with resistivity, i.e., the Sweet-Parker type current sheet, whose aspect ratio is $a/L\sim S_L^{-1/2}$ \citep{sweet1958electromagnetic,parker1957sweet}, the maximum growth rate of tearing is $\gamma \tau_L \sim S_L^{1/4}$ \citep{tajima2018plasma,loureiro2007instability}. The positive power-law index means that the growth rate can be extremely large in the limit $S_L \rightarrow \infty$ \citep{bhattacharjee2009fast,huang2013plasmoid}.

The above analysis leads to a plausible scenario of the explosive energy release of the macroscopic current sheet. Initially, the current sheet is thick with $(a/L) > S_L^{-1/3}$, and thus is stable to tearing mode. Then some external forces gradually build up magnetic energy and result in thinning of the current sheet. Once the current sheet thins to the critical aspect ratio $(a/L) \sim S_L^{-1/3}$, the growth of the tearing mode suddenly becomes very fast, and the current sheet breaks up into many plasmoids and smaller-scale current sheets. This process can happen recursively in the newly formed current sheets and dissipates the magnetic energy rapidly \citep{shibata2001plasmoid,tenerani2015magnetic,landi2015resistive,papini2019fast}, until it is terminated due to the decrease of Lundquist number \citep{shi2018marginal} or the ion kinetic effect \citep{shi2019fast,bora2021evolution}. Thus, tearing instability is an important and fundamental mechanism that facilitates fast reconnection in the large-scale current sheets. Consequently, it is important to thoroughly study it under different configurations. Recent progresses on this topic include the calculations of its linear growth rate with viscosity \citep{tenerani2015tearing}, different background magnetic field profiles \citep{pucci2018onset}, Hall effect \citep{pucci2017fast}, guide field \citep{shi2020oblique}, ion-neutral collisions \citep{pucci2020tearing}, and normal component of magnetic field \citep{shi2021stability}. 

In space plasma, current sheets are frequently accompanied by plasma flows. In the dayside magnetosheath, reconnection events are often observed within highly turbulent plasma \citep[e.g.][]{huang2016mms}, and so for the solar wind \citep[e.g.][]{osman2014magnetic}. As a result, the reconnecting current sheets are likely to be affected by plasma flows of all directions. Plasma flows are also detected in the pre-flare corona \citep[e.g.][]{wallace2010pre} and nightside magnetotail current sheet \citep[e.g.][]{lane2021dynamics}. At the tip of the helmet streamer where the heliospheric current sheet forms, growth of tearing instability accompanied by an outward propagating solar wind stream is observed in MHD simulations \citep{reville2020tearing,reville2022flux}. Thus, study of how the tearing mode instability is modified by plasma flows is necessary. Many works have been conducted on the effect of a shear flow, i.e., flow parallel to the shear magnetic field \citep[e.g.][]{hofman1975resistive,paris1983influence,einaudi1986resistive,chen1990resistive,ofman1991resistive,paris1993effects,dahlburg1997evolution,chen1997tearing,faganello2010collisionless}. For example, \citet{chen1997tearing}, through 2D MHD simulations, show that a sub-Alfv\'enic shear flow stabilizes tearing mode, and with a super-Alfv\'enic shear flow the instability is dominated by Kelvin-Helmholtz mode. This result is confirmed by a recent work \citep{shi2021stability} that calculates the linear instability growth rate by a boundary-value-problem approach. They also show that when the flow is exactly Alfv\'enic, the current sheet is extremely stable and the perturbation only grows at the rate of diffusion. Compared with the shear flow, the case where a plasma jet exists at the center of the current sheet is more complicated, because the jet itself is susceptible to two types of streaming instabilities, i.e., the sausage (varicose) mode and the kink (sinuous) mode. \citet{wang1988streaming}, using an initial value solver of compressible MHD equations, show that a super-Alfv\'enic plasma jet can increase the growth rate of the tearing mode. Subsequent works \citep{wang1988mechanism,lee1988streaming} show that under this type of configuration both the kink mode and sausage mode exist and the sausage mode mixes with the tearing mode in the presence of resistivity. 2D MHD \citep{bettarini2006tearing} and Hall-MHD simulations \citep{hoshino2015generation} confirm these early results.

In this study, we carry out a comprehensive investigation of the stability problem of the current sheet with a plasma jet in the framework of linear incompressible MHD, using an eigenvalue problem solver. We examine both the streaming sausage mode and streaming kink mode with and without resistivity. We derive the controlling equation set under a generalized configuration such that the plasma jet can have arbitrary angle with respect to the magnetic field and a finite guide field is allowed. The paper is organized as follows. In section \ref{sec:equilibrium}, we describe the background fields used in this study. In section \ref{sec:equation_perturb}, we derive the equation set for the perturbation field. In section \ref{sec:results} we present the detailed results of our calculation. In section \ref{sec:summary} we summarize the results and discuss the possible applications of the results to space plasma.

\section{Equilibrium and background fields}\label{sec:equilibrium}
We start from the resistive-MHD equation set
\begin{subequations}
\begin{equation}
    \frac{\partial \rho}{\partial t} + \nabla \cdot \left( \rho \mathbf{V} \right) = 0
\end{equation}
\begin{equation}
   \rho \frac{\partial \mathbf{V}}{\partial t} + \rho \mathbf{V} \cdot \nabla \mathbf{V} = - \nabla P + \mathbf{J}\times \mathbf{B}
\end{equation}
\begin{equation}
    \frac{\partial \mathbf{B}}{\partial t} = \nabla \times \left( \mathbf{V} \times \mathbf{B} \right) + \frac{1}{S} \nabla^2 \mathbf{B}
\end{equation}
\begin{equation}
    \frac{\partial P}{\partial t} + \mathbf{V} \cdot \nabla P + \kappa \left(\nabla \cdot \mathbf{V}\right) P = 0
\end{equation}
\end{subequations}
where $\rho,\mathbf{V},\mathbf{B},P$ are density, velocity, magnetic field, and pressure, $\kappa$ is the adiabatic index, and $S$ is the Lundquist number. Incompressiblility is assumed throughout the study, i.e., $\rho(t,\mathbf{x}) \equiv \rho_0$ is a constant. In a generalized configuration, the background magnetic field and velocity consist of both $x$ and $z$ components but are functions of $y$ only:
\begin{equation}\label{eq:general_form_of_B_V}
    \mathbf{B} = B_x(y) \hat{e}_x + B_z(y) \hat{e}_z, \quad \mathbf{V} = V_x(y) \hat{e}_x + V_z(y) \hat{e}_z.
\end{equation}
Consequently, the momentum convection term and the magnetic tension force are both zero: $\mathbf{V}\cdot \nabla \mathbf{V} \equiv 0, \,\mathbf{B}\cdot \nabla \mathbf{B} \equiv 0$,  and the divergences of $\mathbf{V}$ and $\mathbf{B}$ are also zero. The zeroth-order scalar pressure ensures the pressure balance
\begin{equation}
    P(y) = P^T - \frac{B^2(y)}{2\mu_0},
\end{equation}
where $P^T$ is the uniform total pressure. With the above configuration, the background field is in equilibrium without resistivity, but will diffuse with a finite resistivity. However, in most of the space and laboratory plasmas, the resistivity is extremely small, thus the diffusion time is much longer than the growth time of instabilities of interest. Therefore, we are able to neglect the diffusion of background fields.

In this study, we adopt the Harris type current sheet model for the magnetic field such that $B_x(y)=B_0 \tanh(y/a)$. We also allow a uniform guide field $B_z(y) = B_g$. We assume the velocity is of the following form:
\begin{equation}
    \mathbf{V} = V(y) \left( \cos (\alpha) \, \hat{e}_x + \sin (\alpha) \, \hat{e}_z \right)
\end{equation}
i.e., the jet rotates from the $x$ direction by an angle $\alpha$ (Figure \ref{fig:coordinate_system}a). We adopt a flow function:
\begin{equation}
    V(y) = V_0  \, \mathrm{sech}^2 \left(\frac{y}{d} \right),
\end{equation}
where $d$ is the half-thickness of the jet, and $V_0$ is the flow speed at the center of the current sheet ($y=0$) and is a variable parameter. Throughout the study, we fix $d=a$, i.e., the width of the jet is the same as the width of the current sheet. The profile of $V(y)$ is plotted in Figure (\ref{fig:coordinate_system}b) together with the profile of the $x$-component of the magnetic field $B(y)$.

\section{Equation set for the perturbations}\label{sec:equation_perturb}

\begin{figure}
    \centering
    \includegraphics[width=1\textwidth]{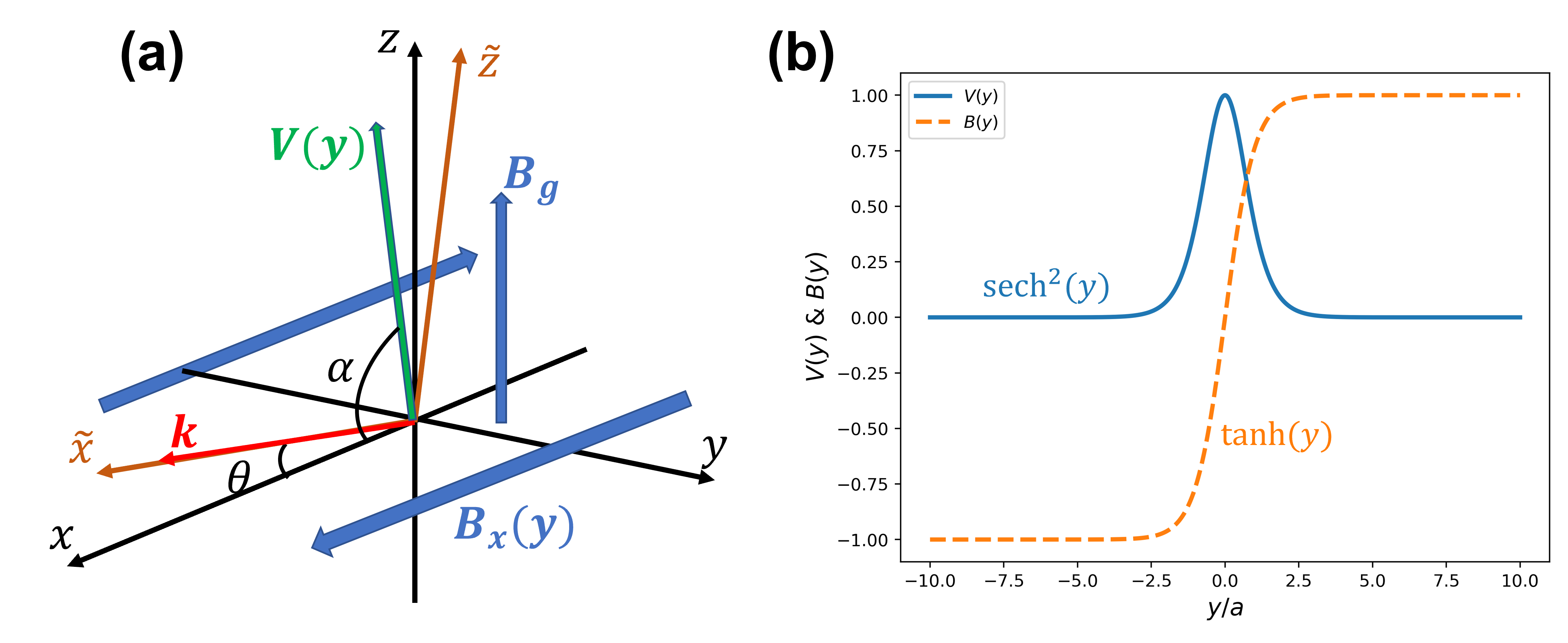}
    \caption{(a) Coordinate systems used in this study and the background fields. Coordinate system $\xt-\zt$ is $x-z$ rotated by an angle $\theta$ with respect to $y$ axis, so that $\xt$ is parallel to the wave vector $\mathbf{k}$. (b) $y$-profiles of the background flow $V(y)$ (blue solid) and the $x$-component of the magnetic field $B(y)$ (orange dashed) used in this study.}
    \label{fig:coordinate_system}
\end{figure}

For the perturbations, we consider a Fourier mode whose growth rate $\gamma$ is a complex number and wave vector $\mathbf{k}$ is in the $x-z$ plane with an arbitrary angle $\theta$ with respect to the $x$ direction (Figure \ref{fig:coordinate_system}a):
\begin{equation}
    \mathbf{k} = k \cos (\theta) \, \hat{e}_x + k \sin (\theta) \, \hat{e}_z.
\end{equation} 
Hence, the perturbation fields have the form:
\begin{equation}
    \left( \begin{array}{c}
         \mathbf{u}(t,\mathbf{x}) \\
         \mathbf{b}(t,\mathbf{x}) \\
         p(t,\mathbf{x})
    \end{array} \right) = \left( \begin{array}{c}
         \mathbf{u} (y) \\
         \mathbf{b} (y) \\
         p (y)
    \end{array} \right)  \exp(\gamma t +i \mathbf{k} \cdot \mathbf{x})
\end{equation}
For simplicity, one can rotate the $x-z$ coordinates with respect to the $y$-axis by angle $\theta$ and get a new coordinate system $\xt-\zt$ (Figure \ref{fig:coordinate_system}a) such that $\mathbf{k} = k \hat{e}_{\xt}$, and thus there is $\partial_{\zt}\equiv 0$. The background magnetic field and velocity can then be written as $\mathbf{B} = B_{\xt}(y) \, \hat{e}_{\xt} + B_{\zt}(y) \, \hat{e}_{\zt}$ and $\mathbf{V} = V_{\xt}(y) \, \hat{e}_{\xt} + V_{\zt}(y) \, \hat{e}_{\zt}$ after projection to the new coordinate system.

The next step is to derive a closed linear equation set for the eigenvalue problem. We start from the linearized momentum equation (with uniform density $\rho$)
\begin{equation}
    \gamma \mathbf{u} + \mathbf{V} \cdot \nabla \mathbf{u} + \mathbf{u} \cdot \nabla \mathbf{V} = - \frac{1}{\rho} \nabla p_{1}^T + \left(\mathbf{B} \cdot \nabla \mathbf{b} + \mathbf{b} \cdot \nabla \mathbf{B}\right),
\end{equation}
where we have normalized the magnetic field by $\sqrt{\mu_0\rho}$ so that it is in the unit of speed. To get rid of the 1st-order pressure, we can take the curl of the equation and get
\begin{equation}
    \gamma \nabla \times \mathbf{u} + \nabla \times \left( \mathbf{V} \cdot \nabla \mathbf{u} + \mathbf{u} \cdot \nabla \mathbf{V}\right) = \nabla \times \left(\mathbf{B} \cdot \nabla \mathbf{b} + \mathbf{b} \cdot \nabla \mathbf{B}\right).
\end{equation}
Using $\nabla \cdot \mathbf{u} = 0$ and $\nabla \cdot \mathbf{b} = 0$ to eliminate $u_{\tilde{x}}$ and $b_{\tilde{x}}$, the $\tilde{z}$ component of the above equation can be re-arranged in the following form:
\begin{equation}\label{app_eq:uy}
    \gamma \left( u_y^{\prime\prime} - k^2 u_y \right) + ik \left[ V_{\xt} \left(u_y^{\prime\prime} - k^2 u_y \right) - V_{\xt}^{\prime\prime} u_y \right] = ik \left[ B_{\xt} \left(b_y^{\prime\prime} - k^2 b_y \right) - B_{\xt}^{\prime\prime} b_y \right]
\end{equation}
where the prime indicates $\partial_y$. We note that the above equation contains only the $y$-component of $\mathbf{u}$ and $\mathbf{b}$. The linearized induction equation is 
\begin{equation}
    \gamma \mathbf{b} = \mathbf{b} \cdot \nabla \mathbf{V} - \mathbf{V} \cdot \nabla \mathbf{b} + \mathbf{B} \cdot \nabla \mathbf{u} - \mathbf{u} \cdot \nabla \mathbf{B} + \frac{1}{S} \nabla^2 \mathbf{b}
\end{equation}
where $S = a V_{A}/\eta$ is defined with the half-thickness of the current sheet $a$ and the upstream Alfv\'en speed $V_A=B_0/\sqrt{\mu_0 \rho}$. The $y$-component of the linearized induction equation is
\begin{equation}\label{app_eq:by}
    \gamma b_y = - ikV_{\xt} b_y + ik B_{\xt} u_y + \frac{1}{S} \left(b_y^{\prime\prime} - k^2 b_y \right)
\end{equation}
which also contains only the $y$-component of $\mathbf{u}$ and $\mathbf{b}$. Thus, equations (\ref{app_eq:uy}) and (\ref{app_eq:by}) form a closed equation set for $u_y$ and $b_y$:
\begin{subequations}\label{eq:u_b_general}
\begin{equation}\label{eq:u_general}
    \gamma \left( u_y^{\prime\prime} - k^2 u_y \right) + ik \left[ V_{\xt} \left(u_y^{\prime\prime} - k^2 u_y \right) - V_{\xt}^{\prime\prime} u_y \right] = k \left[ B_{\xt} \left(b_y^{\prime\prime} - k^2 b_y \right) - B_{\xt}^{\prime\prime} b_y \right]
\end{equation}
\begin{equation}\label{eq:b_general}
     \gamma b_y = - ikV_{\xt} b_y - k B_{\xt} u_y + \frac{1}{S} \left(b_y^{\prime\prime} - k^2 b_y \right)
\end{equation}
\end{subequations}
Here we have assimilated a $\pi/2$ phase difference between $u_y$ and $b_y$, i.e., we have replaced $ib_y$ with $b_y$. %In this way, the equation set becomes pure real without flow.

\begin{figure}
    \centering
    \includegraphics[width=\textwidth]{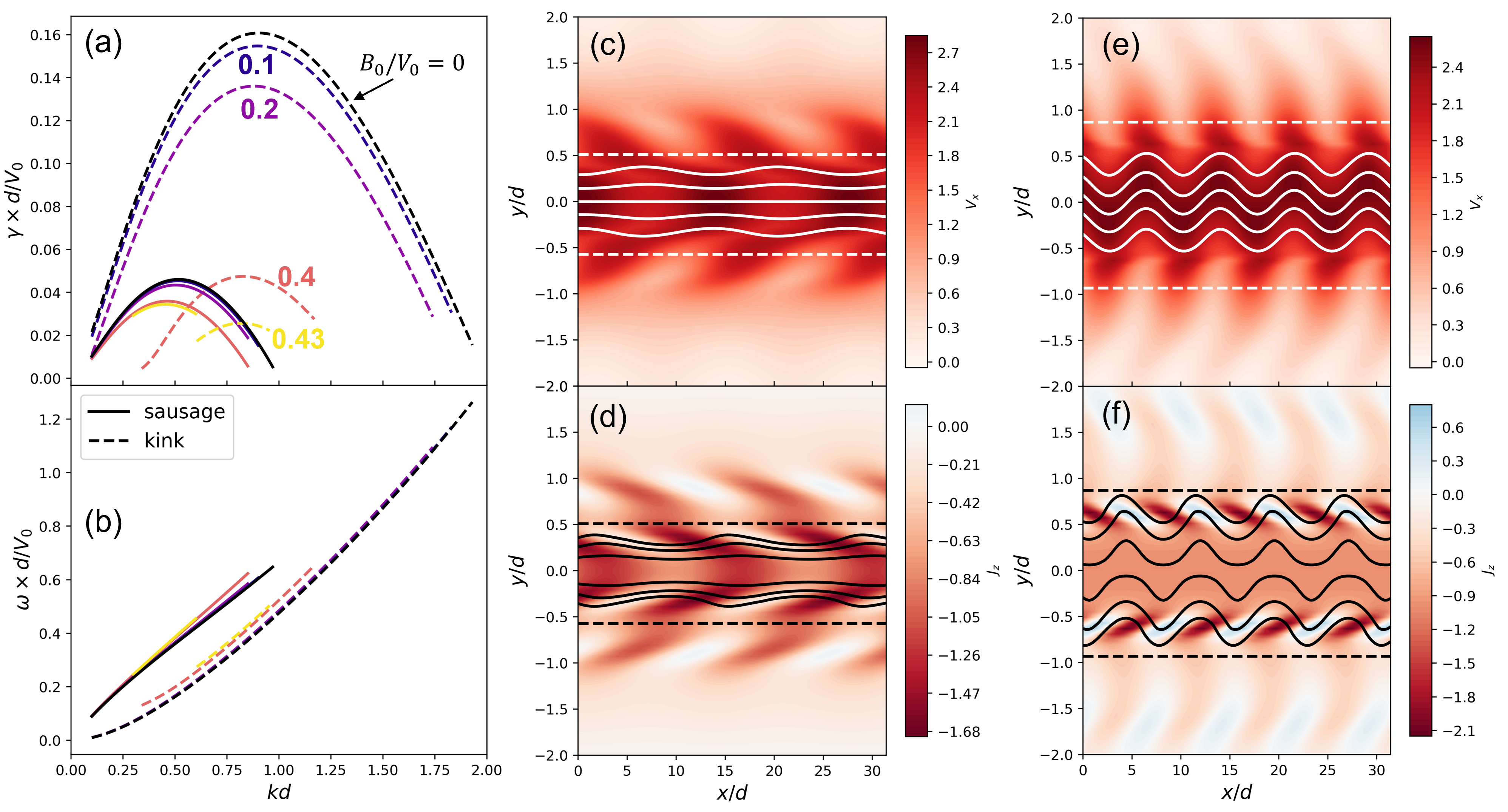}
    \caption{ Streaming instabilities of a plasma jet inside a current sheet. The jet and the wave vector are both parallel to the magnetic field, and there is no guide field. (a,b) Growth rate ($\gamma$) and oscillation frequency ($\omega$) as functions of wavenumber $k$ for the sausage mode (solid lines) and the kink mode (dashed lines) with different magnetic field and jet speed ratios $B_0/V_0$. Black lines represent the non-magneto fluid case ($B_0/V_0 = 0$). Here, the wavenumber is normalized by the half-thickness of the jet, which is equal to the half-thickness of the current sheet, and the growth rate and frequency are normalized to $d/V_0$. Panels (c) and (d) are plotted based on the sausage mode with $B_0/V_0=0.4$ and $kd=0.46$ (the fastest growing mode). (c) 2D profiles of $V_x$, solid lines are the streamlines, and the two dashed lines mark the resonance surfaces where $\omega = k V(y)$. (d) 2D profiles of $J_z$ (out-of-plane current density), solid lines are the magnetic field lines, and the two dashed lines mark the resonance surfaces. Panels (e) and (f) are similar with panels (c) and (d) but for the kink mode with $B_0/V_0=0.4$ and $kd=0.83$ (the fastest growing mode). We note that in panels (c)-(f) all the physical quantities are the sums of the linear eigenfunctions and the background fields.
    %For sausage mode: $k_m=0.52$, $\gamma_m=0.046$. For kink mode: $k_m=0.90$, $\gamma_m=0.16$ 
    }
    \label{fig:nonresistive}
\end{figure}

One can immediately find that, if $V_{\xt} = 0$, i.e., if $\mathbf{k} \cdot \mathbf{V} = 0$, the system is purely determined by $B_{\xt}$, similar to the classic tearing case, and the growth rate is purely real. As an example, consider an anti-parallel magnetic field $\mathbf{B}=B_x(y) \hat{e}_x$ and an out-of-plane flow $\mathbf{V}=V_z(y) \hat{e}_z$. In this case, if $\mathbf{k}=k \hat{e}_x$, the system reduces to the classic tearing, i.e., the flow has no effect on the solution. But if the wave vector is not along $x$ and has a finite $z$-component, the flow will alter the growth rate and introduce an oscillation (a non-zero frequency) to the solution. Similarly, if $\mathbf{k} \cdot \mathbf{B} = 0$, the system is determined purely by the flow $V_{\xt}$ and there are only stream-induced instabilities. 

As a final remark, we note that Equation (\ref{eq:u_b_general}) is in generalized form, and works for any functions $\mathbf{V}(y)$ and $\mathbf{B}(y)$ once they have the form of equation (\ref{eq:general_form_of_B_V}) and the system is incompressible. 

\section{Results}\label{sec:results}
\begin{figure}
    \centering
    \includegraphics[width=\textwidth]{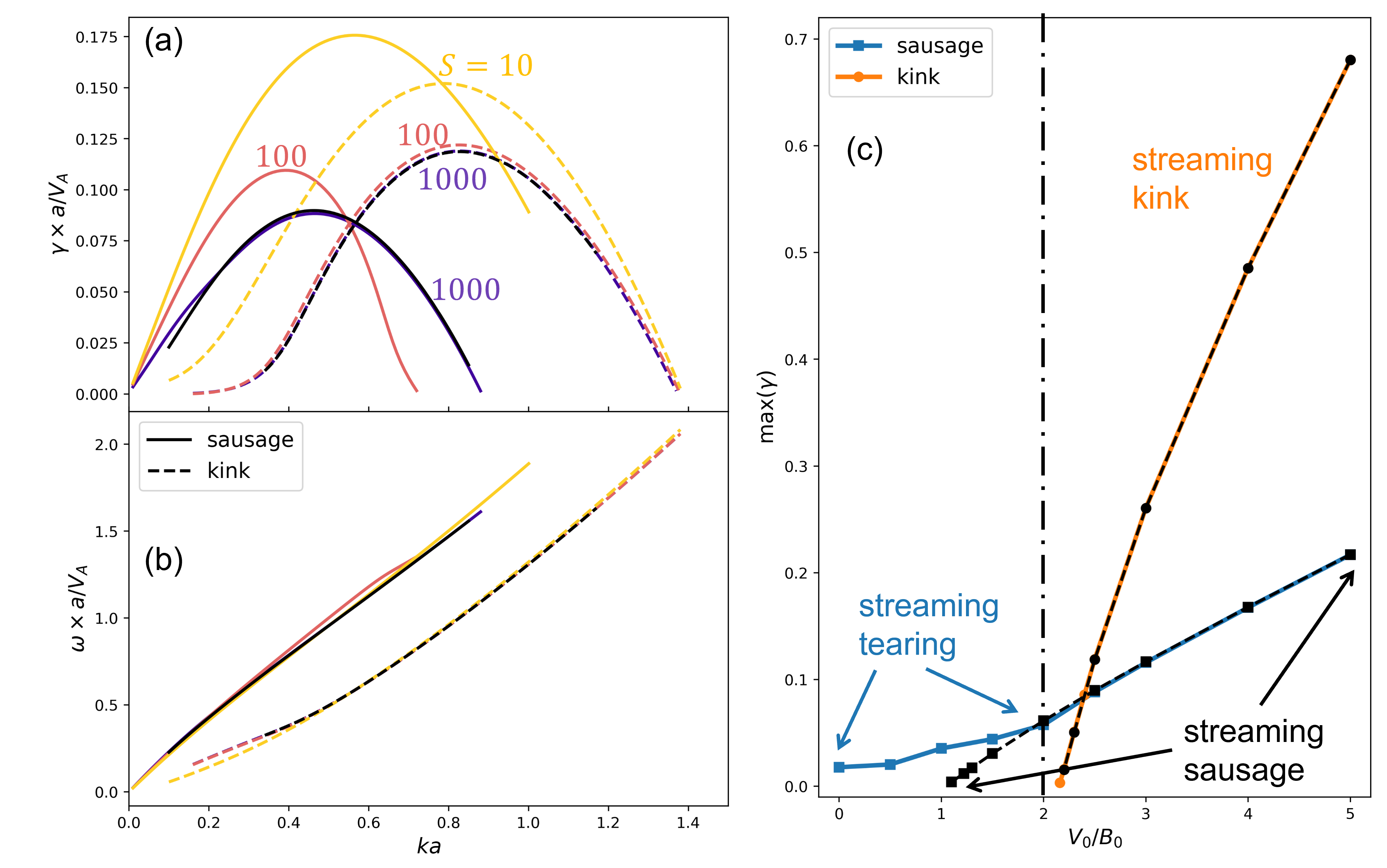}
    \caption{(a, b) Dispersion relation $\gamma(k)$ \& $\omega(k)$ for sausage mode (solid lines) and kink mode (dashed lines) with $V_0/B_0=2.5$. Colors of the curves correspond to the Lundquist numbers, such that yellow is $S=10$, light purple is $S=100$, and dark purple is $S=1000$. Black curves are non-resistive cases ($S\rightarrow \infty$). Different from Figure \ref{fig:nonresistive}, here $\gamma$ and $\omega$ are normalized by the Alfv\'en crossing time $a/V_A$ (or $a/B_0$). (c) maximum growth rate max($\gamma(k)$) as a function of $V_0/B_0$. Blue and orange curves are the sausage and kink modes respectively, with $S=1000$. The two black dashed curves are the non-resistive cases ($S\rightarrow \infty$).}
    \label{fig:disp_rel_diff_S_max_gamma_with_V0}
\end{figure}

Equation (\ref{eq:u_b_general}) is a boundary-value-problem (BVP) with the boundary conditions $u_y(y\rightarrow \pm \infty)=0, b_y(y\rightarrow \pm \infty)=0$. Far from the center of the current sheet ($y\rightarrow \pm \infty$), the derivatives of the background fields reduce to zero. Consequently, one can see that $u_y,b_y \propto \exp(-k|y|)$ satisfy equation (\ref{eq:u_general}), and equation (\ref{eq:b_general}) just gives the ratio $u_y/b_y$ at $y\rightarrow \pm\infty$. In this study, we use the numerical BVP solver implemented in the Python package SciPy \citep{virtanen2020scipy} to solve equation (\ref{eq:u_b_general}). In practice, we set the boundaries at $y=\pm 15a$, which are large enough to acquire accurate solutions. 

Before solving the equation set, we need to define the parities of $u_y$ and $b_y$ first. Given that $V_{\xt}(y)$ is an even function and $B_{\xt}(y)$ is an odd function, one can see from equation (\ref{eq:u_b_general}) that there are two possible combinations of the parities of $u_y$ and $b_y$: (1) $u_y(y)$ is odd and $b_{y}(y)$ is even, (2) $u_y(y)$ is even and $b_{y}(y)$ is odd. The first case is the so-called ``sausage'' (varicose) mode, which leads to the formation of a chain of blobs and plasmoids (if resistivity is non-zero). Tearing instability is categorized to the sausage mode. The second case is the ``kink'' (sinuous) mode, which results in wavy distortion of the current sheet and stream. In solving the problem, we carefully search for both of the two modes. 

% The growth rate of the kink mode may be comparable to the sausage mode when the stream speed along $x$ is super-Alfv\'enic \citep{lee1988streaming}, but it is almost independent on the resistivity.

\subsection{Instabilities of a plasma jet inside a non-resistive current sheet}\label{sec:result_no_resis}

In this section, we analyze the simplest 2D case, where the jet, magnetic field, and wave vector are all aligned along the $x$ direction, i.e., there is no out-of-plane flow or guide field. Besides, the current sheet is non-resistive. We will show how the pure streaming instabilities are modified by the interaction between the jet and the current sheet.

In panels (a) \& (b) of Figure \ref{fig:nonresistive}, we show the dispersion relations $\gamma(k)$ and $\omega(k)$ for the sausage mode (solid lines) and kink mode (dashed lines). The wavenumber is normalized to $d$, and $\gamma$ \& $\omega$ are normalized to $V_0/d$. The lines are color-coded with the $B_0/V_0$ ratio, as marked in panel (a). Both modes have larger growth rate with smaller magnetic field, as it is well-known that the magnetic field suppresses the stream-shear instability. The kink mode has larger growth rate than the sausage mode for $B_0/V_0 \lesssim 0.4$. But when $B_0/V_0 \gtrsim 0.4$, the sausage mode becomes more unstable than the kink mode, i.e., the magnetic field stabilizes the kink mode more effectively. The wavenumber of the most unstable kink mode is roughly twice the wavenumber of the most unstable sausage mode. In panels (c) \& (d), we show 2D profiles of $V_x$ and $J_z$ respectively for the sausage mode with $B_0/V_0=0.4$ and $kd=0.46$, which is the fastest growing mode for $B_0/V_0=0.4$. Here, the 2D profiles are calculated by summing the background fields and the solved eigenfunctions of the perturbations. The solid curves in panels (c) \& (d) are the streamlines and the magnetic field lines respectively, and one can see the formation of a chain of sausage-like structures in both the velocity and the magnetic field. The dashed lines in the two panels mark the resonance surfaces where $\omega = k V(y)$ and the fields undergo sharp transitions. Panels (e) \& (f) are similar to (c) \& (d) but for kink mode with $B_0/V_0=0.4$ and $kd=0.83$, which is the fastest growing mode for $B_0/V_0=0.4$. We can see the growth of kink mode deforms the jet and the magnetic field in a sinuous way.

Here we must point out that, in the non-resistive case, growth of the streaming sausage mode does not result in reconnection of the magnetic field since the system is ideal-MHD. The sausage-like structures are not plasmoids but are merely anti-phased deformation of the magnetic field lines on the two sides of the current sheet (panel (d) of Figure \ref{fig:nonresistive}). Actually, from equation (\ref{eq:b_general}), one can see that $b_y(y=0) = 0$ in the limit $S\rightarrow \infty$.

%Cannot find solutions when $V_0/B_0 \lesssim 2.5$ for kink mode and $V_0/B_0 \lesssim 1.5$ for sausage mode. Because there are two resonance surfaces $\omega = k V(y)$ at which some denominators are almost zero.

\begin{figure}
    \centering
    \includegraphics[width=\textwidth]{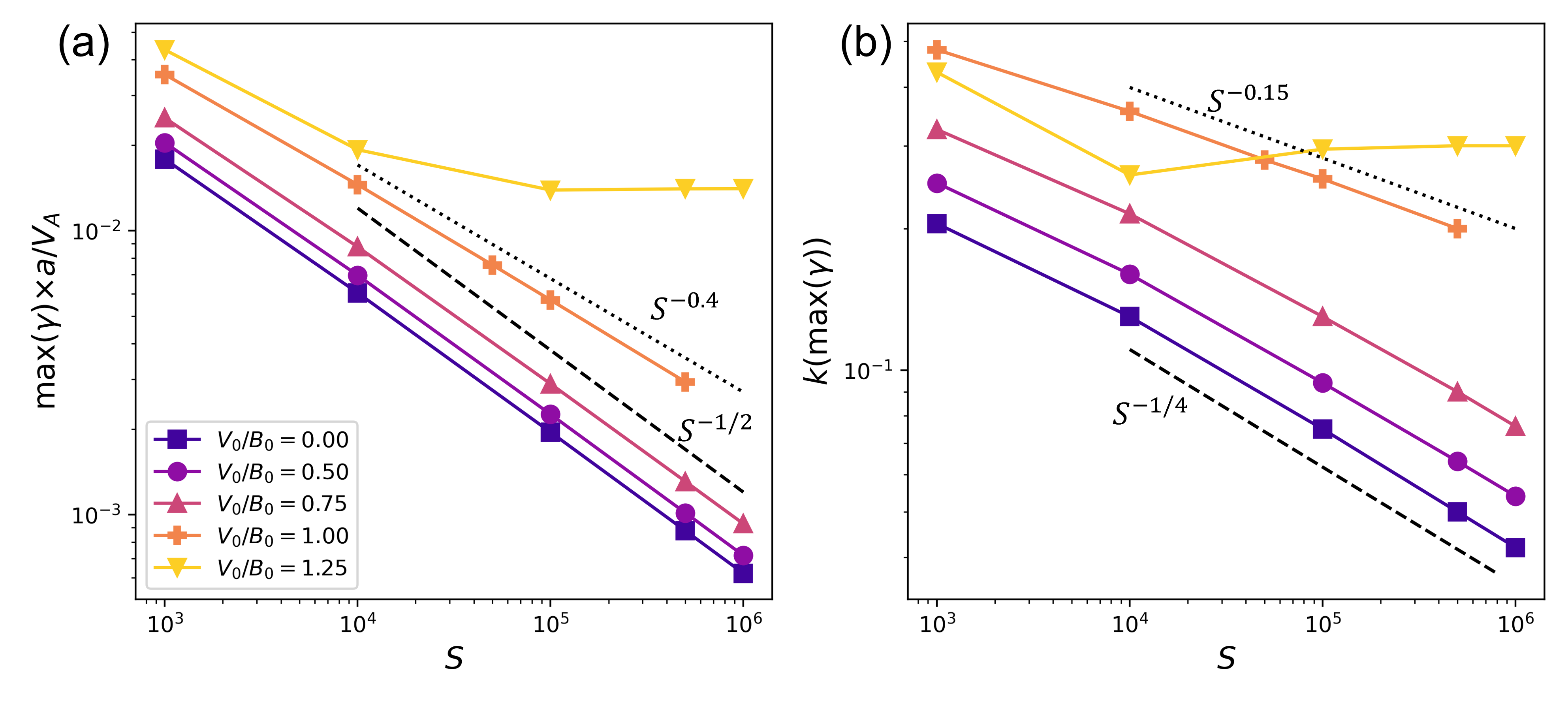}
    \caption{(a) Maximum growth rate of the sausage mode as a function of the Lundquist number $S$ for different $V_0/B_0$. The black dashed line shows $\gamma \propto S^{-1/2}$ and the black dotted line shows $\gamma \propto S^{-0.4}$ for reference. (b) Corresponding wavenumber of the most unstable mode as a function of $S$. The black dashed line shows $k \propto S^{-1/4}$ and the black dotted line shows $k \propto S^{-0.15}$ for reference.}
    \label{fig:max_gamma_with_S_diff_V0}
\end{figure}

\subsection{Instabilities of a plasma jet inside a resistive current sheet}\label{sec:result_resis_streaming}
Based on the results of the previous section, we explore how a finite resistivity will modify the streaming instabilities and how the tearing mode couples with the streaming sausage mode.

Panels (a) \& (b) of Figure \ref{fig:disp_rel_diff_S_max_gamma_with_V0} show the dispersion relation $\gamma(k)$ and $\omega(k)$ for the sausage mode (solid lines) and kink mode (dashed lines) with $V_0/B_0=2.5$ and different Lundquist numbers. Different from Figure \ref{fig:nonresistive}, hereinafter we normalize $\gamma$ \& $\omega$ by the Alfv\'en crossing time $a/V_A$, i.e. $a/B_0$, which is conventional for the analysis of tearing mode. Although not very significantly at this $V_0/B_0$ ratio, resistivity enhances the maximum growth rates of both modes. It is clear that kink mode is less affected by resistivity compared with the sausage mode. This is a reasonable result because the streaming sausage mode is expected to couple with the tearing mode whose growth rate is determined by the resistivity. To better illustrate this point, panel (c) shows the maximum growth rates max($\gamma$) as functions of the ratio $V_0/B_0$ for the sausage and kink modes. The blue line is the sausage mode with $S=1000$, and the orange line is the kink mode with $S=1000$. The two black lines are the cases without resistivity ($S\rightarrow \infty$). As already shown in section \ref{sec:result_no_resis}, in the non-resistive case, the kink mode (black squares) is more unstable than the sausage mode (black circles) with large $V_0/B_0$ but its growth rate decreases fast toward zero as $V_0/B_0$ approaches $\sim 2$ from above. The growth rate of sausage mode decreases to zero at $V_0/B_0 \sim 1$. With $S=1000$, the maximum growth rate of the kink mode almost does not change from the non-resistive case. For the sausage mode, the growth rate does not change for $V_0/B_0 \geq 2$ but becomes larger than the non-resistive case for $V_0/B_0 <2$. Especially, even for $V_0/B_0 < 1$, its maximum growth rate is larger than zero. This is because of the transition of the streaming sausage mode to the tearing mode at small $V_0/B_0$ ratio. Thus, in the finite (but not too large) resistivity case, we can roughly divide the sausage mode into two regimes according to $V_0/B_0$. For $V_0/B_0 \gtrsim 2$, the mode is almost dominated by the jet, hence it is the ``streaming sausage'' mode. For $V_0/B_0 \lesssim 2$, especially for $V_0/B_0 \lesssim  1$, the mode becomes heavily affected by the resistivity, thus it is the ``streaming tearing'' mode. Similar result was obtained by \citet{wang1988streaming}, who showed that there is a sharp increase of the maximum growth rate of the sausage mode at $V_0/B_0 \approx 1.2$, implying a transition of the tearing mode to streaming mode.

%We find that a finite resistivity increases the growth rate of kink mode. Also, with larger and larger resistivity (smaller $S$), we can decrease $V_0$. For $S=10^{-4}$, $V_0$ can decrease to below 0.1. This is natural, because as $S\rightarrow 0$, we expect $b_y \rightarrow 0$. In this case, the background magnetic field has almost no effect on the system and the system reduces to the pure streaming kink instability. We have verified this point by checking the dispersion relation for $S=10^{-4}$. 

Panel (a) of Figure \ref{fig:max_gamma_with_S_diff_V0} shows how the maximum growth rate of the sausage mode scales with the Lundquist number for different $V_0/B_0$, and panel(b) shows the corresponding wavenumbers. The black dashed and dotted lines in panel (a) show $\gamma \propto S^{-1/2}$ and $\gamma \propto S^{-0.4}$ for references, and those in panel (b) show $k \propto S^{-1/4}$ and $k\propto S^{-0.15}$ for references. Classic tearing mode theory \citep{furth1963finite,coppi1966resistive} shows that the maximum growth rate and corresponding wavenumber have the scaling relations $\gamma \propto S^{-1/2}$ and $k \propto S^{-1/4}$ in the limit of large $S$. This is confirmed by the result for $V_0/B_0 = 0$ in Figure \ref{fig:max_gamma_with_S_diff_V0}. As $V_0/B_0$ increases, both the maximum growth rate and the wavenumber increase, but the slopes of these lines remain unchanged for $V_0/B_0 <1$, implying that the mode is still tearing-like. But at $V_0/B_0 = 1$, the scaling relations change such that the two lines are less steep, meaning that the dependence of the instability on the resistivity becomes weaker. At $V_0/B_0=1.25$, the two lines are flat for $S \geq 10^5$, indicating that the instability becomes weakly dependent on resistivity and start to transit to pure streaming sausage mode.

\begin{figure}
    \centering
    \includegraphics[width=\textwidth]{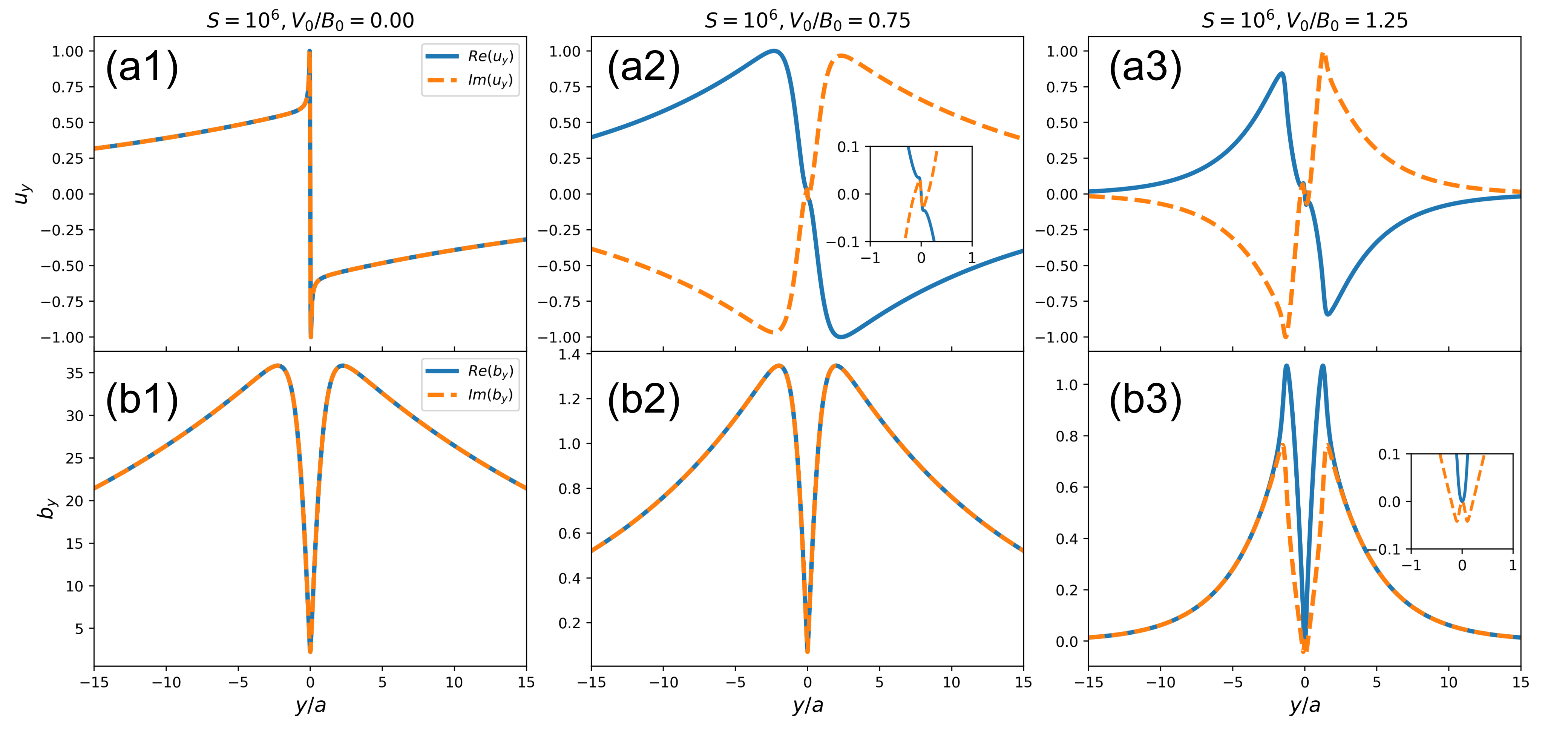}
    \caption{Eigenfunctions $u_y$ (row (a)) and $b_y$ (row (b)) for the most unstable sausage modes with $S=10^6$ and varying $V_0/B_0$. From left to right columns are $V_0/B_0=$0, 0.75, and 1.25 respectively. In each panel, the solid and dashed curves are the real and imaginary parts of the eigenfunctions. In panels (a2) and (b3), the embedded plots show the blow-ups of the eigenfunctions.}
    \label{fig:eigenfunctions_diff_V0}
\end{figure}

In Figure \ref{fig:eigenfunctions_diff_V0}, we show the solved eigenfunctions $u_y$ (row (a)) and $b_y$ (row (b)) for the most unstable sausage modes with $S=10^6$. From left to right columns are $V_0/B_0=$0, 0.75, and 1.25 respectively. In each panel, the solid curve is the real part, and the dashed curve is the imaginary part of the function. Panels (a1) and (b1) correspond to the classic tearing case, where the solution shows a very thin singular layer, or ``inner'' layer, around $y=0$, which is dominated by the resistivity. Outside the singular layer, the solution is determined by the non-resistive parts of equation (\ref{eq:u_b_general}). As the ratio $V_0/B_0$ increases (middle and right columns), the outer solution is altered by the jet, while the inner singular layer persists, as can be seen in the embedded blow-up plots. In addition, as $V_0/B_0$ increases, the relative amplitude of $b_y$ and $u_y$ drops. For the pure tearing mode, $|b_y| \gg |u_y|$, implying the mode is magnetic field dominated. As the mode transits to streaming sausage mode, magnetic field perturbation gradually weakens.

\subsection{Jet along the guide field and the oblique tearing mode}\label{sec:oblique_mode}
\begin{figure}
    \centering
    \includegraphics[width=\textwidth]{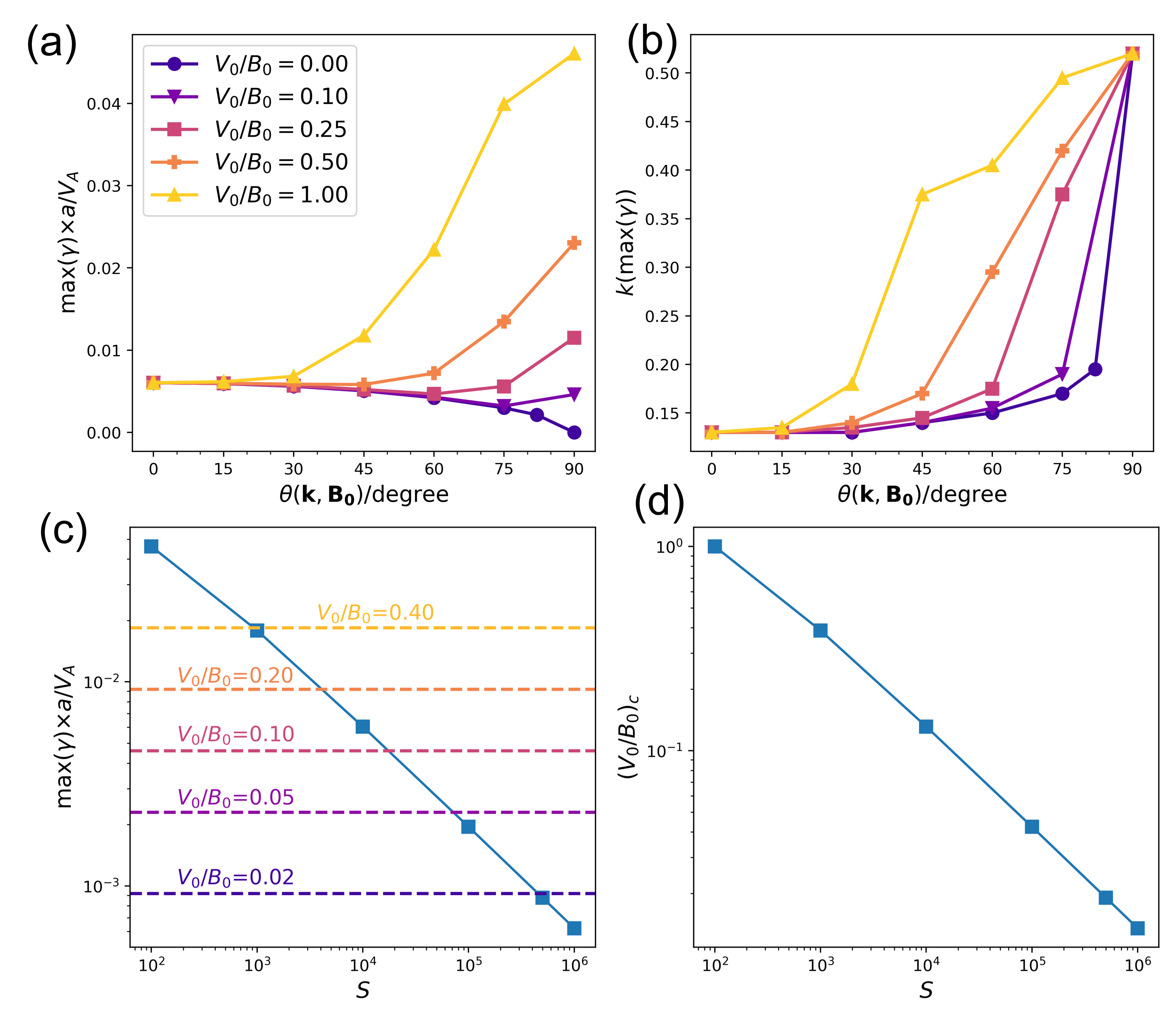}
    \caption{(a) Maximum growth rate of the sausage mode as a function of $\theta$ (angle between $\mathbf{k}$ and $\mathbf{B_0}$), for $\alpha=90^\circ$ (angle between $\mathbf{V_0}$ and $\mathbf{B_0}$), $S=10^4$, $B_g=0$, and varying $V_0/B_0$. (b) The corresponding wavenumbers. (c) Blue curve with square markers: maximum growth rate of pure tearing mode, i.e. $\theta= 0$, as a function of $S$. Horizontal dashed lines mark the maximum growth rate of the pure streaming sausage mode, i.e. $\theta=90^\circ$, with varying $V_0/B_0$. Note that the growth rates of pure streaming modes (modes decoupled from the magnetic field) are independent of $S$. (d) Critical value $(V_0/B_0)_c$, above which the pure streaming sausage mode has larger maximum growth rate than the pure tearing mode, as a function of $S$.}
    \label{fig:max_gamma_with_theta}
\end{figure}
In this section, we only discuss the sausage mode because the kink mode is not affected much by the resistivity and not directly related to the reconnection process. In space environments or laboratory plasma, the jet is not necessarily parallel to the reconnecting magnetic field component, and it is possible that the jet has a finite component along the guide field direction (generally speaking, the $z$ direction no matter whether a guide field exists). In this case, the most unstable mode may be oblique whose wave vector also has a component along the guide field. In this section, we consider the case where the jet is along $z$ axis. As the instability is determined fully by the background fields projected on the wave vector direction, one can imagine that, when we rotate $\mathbf{k}$ from $x$ direction to $z$ direction, the instability transits from pure tearing mode to pure streaming mode.

Panel (a) of Figure \ref{fig:max_gamma_with_theta} shows the maximum growth rate of the sausage mode as a function of $\theta(\mathbf{k},\mathbf{B_0})$, which is the angle between $\mathbf{k}$ and $x$-axis, for $S=10^4$, $B_g=0$, and different $V_0/B_0$ ratios. If there is no jet ($V_0/B_0 = 0$), increasing $\theta$ leads to a monotonic decrease of the maximum growth rate. But as the jet speed increases, there is a turning point from which the maximum growth rate starts to rise. Even for a small ratio $V_0/B_0=0.1$, a turning point exists at large $\theta$($ \approx 75^\circ$). However, in this case the perpendicular ($\theta=90^\circ$) mode still grows slower than the parallel ($\theta=0^\circ$) mode. For large jet speed ($V_0/B_0=$1.00), the curve is monotonically increasing. The turning point is due to the transition from the pure tearing mode to the pure streaming sausage mode as $\mathbf{k}$ rotates. Panel (b) shows the wavenumbers corresponding to the modes shown in panel (a). In general, as $\theta$ increases, the wavenumber also increases, especially for large $V_0/B_0$ ratios, because the most unstable streaming mode has larger wavenumber than the most unstable tearing mode. In panel (c), we plot the maximum growth rate of pure tearing mode as a function of $S$ in blue line with square markers. This curve corresponds to the $\theta=0$ case with $\mathbf{V_0} \parallel \hat{e}_z$. In this panel, the horizontal dashed lines mark the maximum growth rate of the pure streaming sausage mode, i.e. the $\theta=90^\circ$ case, for different $V_0/B_0$ ratios. We note that the growth rates of the pure streaming modes are independent of $S$ because these modes decouple with the magnetic field. Consequently, the growth rate is simply proportional to $V_0/B_0$. From this plot, we see that at any fixed Lundquist number $S$, there is a critical $V_0/B_0$ above which the perpendicular mode (pure streaming sausage) has larger growth rate than the parallel mode (pure tearing). In panel (d) we plot this critical value $(V_0/B_0)_c$ as a function of $S$. One can read that, for example, for $S=10^3$, the critical value is $V_0/B_0\approx 0.4$ while for $S=5\times 10^5$ the critical value is $V_0/B_0\approx 0.02$. For a $V_0/B_0$ that exceeds the critical value, the max$(\gamma)-\theta$ curve (such as those shown in panel (a)) will raise with $\theta$ at some point and eventually reach a value at $\theta=90^\circ$ higher than that at $\theta = 0^\circ$.

\begin{figure}
    \centering
    \includegraphics[width=\textwidth]{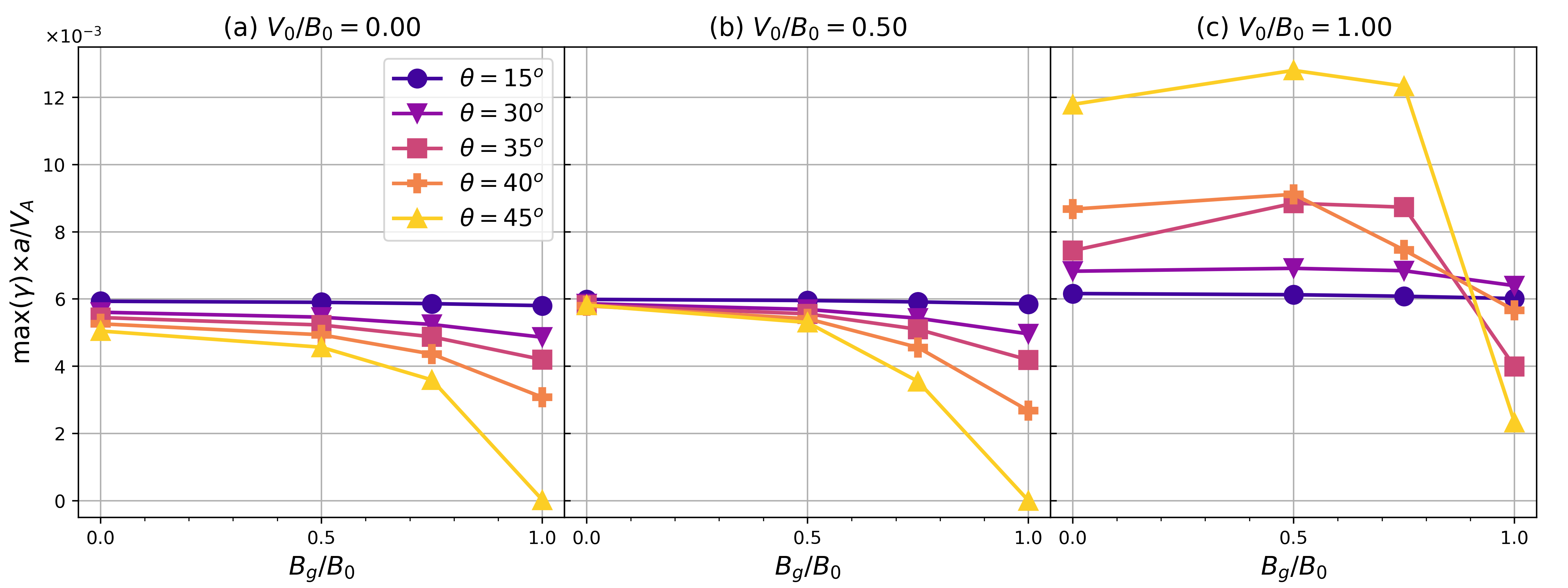}
    \caption{Maximum growth rate of the sausage mode as a function of the guide field strength $B_g/B_0$ for different $\theta$ (angle between $\mathbf{k}$ and $x$-axis). The jet is along the guide field ($\alpha=90^\circ$). The Lundquist number is $S=10^4$. The three panels are results for different flow speeds.}
    \label{fig:max_gamma_with_Bg}
\end{figure}

Then we consider the case with a uniform guide field $B_g$. In Figure \ref{fig:max_gamma_with_Bg}, each panel displays the maximum growth rate of the sausage mode as a function of the guide field strength $B_g/B_0$ for different $\theta$. Panels (a)-(c) correspond to $V_0/B_0=$0, 0.5, and 1.0 respectively. We note that $B_0$ is the asymptotic amplitude of the $x$-component of the magnetic field. For small $\theta$ ($\theta \lesssim 15^\circ$), the maximum growth rate is not significantly modified by either $B_g$ or $V_0$ since $\mathbf{k}$ is quasi-perpendicular to the guide field direction. As $\theta$ increases, for small and intermediate jet speeds ($V_0/B_0 \leq 0.5$), the maximum growth rate drops with the guide field. As already shown by \citet{shi2020oblique}, in the no-flow case, the guide field raises the growth rate only at large-$k$ (the so-called constant-$\psi$) regime, but overall the maximum growth rate of the oblique mode ($\theta$>0) decreases with an increasing guide field strength. However, panel (c) ($V_0/B_0 = 1$) shows a very different result. For small guide field $B_g/B_0 \leq 0.5$, the maximum growth rate increases with $\theta$, similar to the result shown by Figure \ref{fig:max_gamma_with_theta}, because the streaming mode of the jet has larger growth rate than the tearing mode. As $B_g/B_0$ continues to increase ($B_g/B_0=$0.75 and 1.0), max($\gamma$) does not monotonically increase with $\theta$ but may start to drop with $\theta$. Clearly, there is a competition between the jet and guide field. The jet tends to increase max$(\gamma)$ as $\mathbf{k}$ rotates from $x$ direction toward $z$ direction, while the guide field overall plays a counter role but at certain $\theta$ it may raise max$(\gamma)$ first before declining it (see curves for $\theta=30^\circ$, $40^\circ$, and $45^\circ$). Figure \ref{fig:max_gamma_with_theta_different_Bg} displays the maximum growth rate of the sausage mode as a function of $\theta$ in the case $S=10^4$, $V_0/B_0=1$, and $\alpha=90^\circ$, with different $B_g/B_0$ ratios. Without the guide field, max($\gamma$) monotonically increases with $\theta$ as already shown in Figure \ref{fig:max_gamma_with_theta}. But as $B_g/B_0$ increases, peaks appear in the max($\gamma$)-$\theta$ curves, because the guide field effectively stabilizes both the oblique tearing mode and the streaming sausage mode, and hence a finite $B_g$ can significantly decrease the growth rate of the perpendicular mode ($\theta=90^\circ$).

\section{Summary} \label{sec:summary}
\begin{figure}
    \centering
    \includegraphics[width=0.7\textwidth]{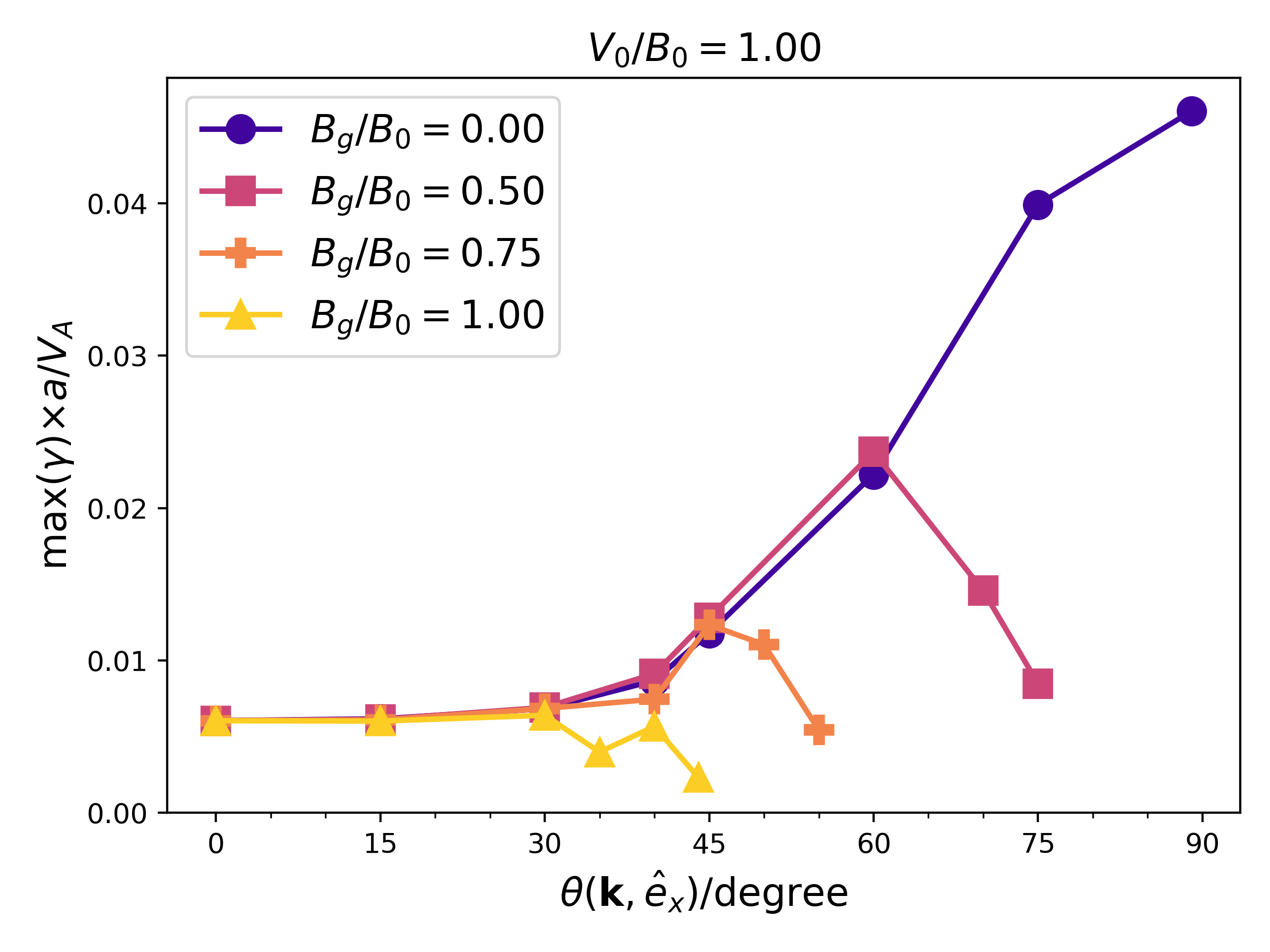}
    \caption{Maximum growth rate of the sausage mode as a function of $\theta$ for $S=10^4$, $V_0/B_0=1$, $\alpha=90^\circ$, and different $B_g/B_0$.}
    \label{fig:max_gamma_with_theta_different_Bg}
\end{figure}

In this study, we adopt a boundary-value-problem solver to study the instabilities inside a current sheet with the presence of a plasma jet. When the jet is collimated with the anti-parallel component of the magnetic field, both of the sausage mode and kink mode can be stabilized by the magnetic field. Without resistivity, the stability thresholds for the kink mode and sausage mode are $V_0/B_0\approx 2$ and $V_0/B_0 \approx 1$ respectively (Figure \ref{fig:disp_rel_diff_S_max_gamma_with_V0}). With a finite resistivity, the streaming sausage mode couples with the tearing mode, but the streaming kink mode is not modified by the resistivity significantly unless the resistivity is very large ($S<100$). Thus, in most of space and laboratory current sheets where $S$ is extremely large, the kink mode can be excited only if the jet speed is large ($V_0/B_0 \gtrsim 2$). For $V_0/B_0 \lesssim 1$, the sausage mode is tearing-like, with a power-law relation between the maximum growth rate and the Lundquist number max$(\gamma)\propto S^{-1/2}$ in the large $S$ limit, same as the tearing mode without flow, while the values of the maximum growth rate increase with the ratio $V_0/B_0$. For $V_0/B_0 \gtrsim 1$, the sausage mode gradually transits to more streaming-like, and the maximum growth rate becomes less dependent on $S$ (Figure \ref{fig:max_gamma_with_S_diff_V0}). In the case of a jet flowing along the direction perpendicular to the anti-parallel component of the magnetic field, our result reveals that, once the jet speed exceeds a threshold which is determined by the Lundquist number, the maximum growth rate of the sausage mode may increase with the angle between the wave vector and the reconnecting magnetic field component (Figure \ref{fig:max_gamma_with_theta}). This is because the mode transits from the pure tearing to pure streaming as the wave vector rotates from the anti-parallel magnetic field direction to the jet direction. Last, the out-of-plane jet combined with a finite guide field leads to a complex behavior of the maximum growth rate of the sausage mode. With certain $V_0/B_0$ and $\theta$ values (panel (c) of Figure \ref{fig:max_gamma_with_Bg} and Figure \ref{fig:max_gamma_with_theta_different_Bg}), the maximum growth rate increases with the guide field strength. But the increase is not very large and is non-monotonically dependent on $\theta$. For example, panel (c) of Figure \ref{fig:max_gamma_with_Bg} shows that the increase in max$(\gamma)$ with $B_g$ from $B_g/B_0=0$ to $B_g/B_0=0.5$ is larger for $\theta=30^\circ$ and 45$^\circ$ than that for $\theta=40^\circ$. More importantly, overall the guide field quenches both the oblique tearing mode and the streaming sausage mode. As a result, increasing $B_g$ will gradually turn the monotonically increasing max$(\gamma)$-$\theta$ curve to a curve that increases at first and then drops (Figure \ref{fig:max_gamma_with_theta_different_Bg}). 

These results indicate that plasma flow plays an important role in destabilizing the current sheets in space and laboratory plasma. A jet whose width is comparable to that of the current sheet and peak speed similar to the upstream Alfv\'en speed can enhance the maximum growth rate of the tearing mode to more than twice of that in the no-flow case (Figure \ref{fig:disp_rel_diff_S_max_gamma_with_V0}). When the jet has a finite component along the direction perpendicular to the anti-parallel component of the magnetic field, even if the component is much smaller than the upstream Alfv\'en speed, the oblique sausage mode ($\theta>0$) may have comparable or even larger growth rate than the parallel sausage mode ($\theta = 0$), and the most unstable mode may be perpendicular ($\mathbf{k}=k \hat{e}_z$) (Figure \ref{fig:max_gamma_with_theta}). The reason is that the out-of-plane flow (along $z$ direction) does not feel the stabilization effect by the magnetic field along $x$, and the growth rate of the pure streaming sausage mode is usually much larger than the pure tearing mode in the large $S$ limit. When the out-of-plane jet and guide field coexist, the most unstable mode may be oblique rather than parallel or perpendicular (Figure \ref{fig:max_gamma_with_theta_different_Bg}). 

We note that several factors which are absent in this study may have non-negligible effects on the analyzed instabilities. Here we assume a uniform density profile and incompressibility. However, compressible MHD simulations show that a nonuniform background plasma density such as in the magnetotail can modify the growth rate of both tearing and streaming modes \citep{hoshino2015generation}. In addition, if Hall effect is included, out-of-plane components of the magnetic field and velocity perturbations are generated even for the parallel mode ($\mathbf{k}=k \hat{e}_x$). Therefore, the out-of-plane jet will modify both the oblique and the parallel modes. Moreover, different widths of the jet and current sheet will change the results \citep{einaudi1986resistive,hoshino2015generation}. As a final remark, it is worth noting that in collisionless regime where the electron inertia is the only mechanism that breaks the frozen-in condition, an out-of-plane plasma jet plays a stabilizing role of the tearing mode even if the mode is parallel ($\mathbf{k}=k \hat{e}_x$) \citep{tassi2014linear}. This is very different from the resistive-MHD regime where the out-of-plane jet only modifies the oblique tearing mode.

Acknowledgements: The author thanks Prof. Marco Velli and Dr. Kun Zhang for very helpful suggestions and comments, and the SciPy team for implementing the boundary-value-solver in Python \citep{virtanen2020scipy}.
The work was supported by NASA HERMES DRIVE Science Center grant No. 80NSSC20K0604.

\bibliographystyle{jpp}
% Note the spaces between the initials

\bibliography{references}

\begin{thebibliography}{60}
\expandafter\ifx\csname natexlab\endcsname\relax\def\natexlab#1{#1}\fi
\def\au#1{#1} \def\ed#1{#1} \def\yr#1{#1}\def\at#1{#1}\def\jt#1{\textit{#1}}
  \def\bt#1{#1}\def\bvol#1{\textbf{#1}} \def\vol#1{#1} \def\pg#1{#1}
  \def\publ#1{#1}\def\arxiv#1{#1}\def\org#1{#1}\def\st#1{\textit{#1}}

\bibitem[Angelopoulos {\em et~al.\/}(2008)Angelopoulos, McFadden, Larson,
  Carlson, Mende, Frey, Phan, Sibeck, Glassmeier, Auster {\em
  et~al.\/}]{angelopoulos2008tail}
{\sc \au{Angelopoulos, Vassilis}, \au{McFadden, James~P}, \au{Larson, Davin},
  \au{Carlson, Charles~W}, \au{Mende, Stephen~B}, \au{Frey, Harald}, \au{Phan,
  Tai}, \au{Sibeck, David~G}, \au{Glassmeier, Karl-Heinz}, \au{Auster, Uli} \&
  \au{others}} \yr{2008}  \at{Tail reconnection triggering substorm onset}.
  \jt{Science}  \bvol{321}~(5891),  \pg{931--935}.

\bibitem[Bettarini {\em et~al.\/}(2006)Bettarini, Landi, Rappazzo, Velli \&
  Opher]{bettarini2006tearing}
{\sc \au{Bettarini, L}, \au{Landi, S}, \au{Rappazzo, FA}, \au{Velli, M} \&
  \au{Opher, M}} \yr{2006}  \at{Tearing and kelvin-helmholtz instabilities in
  the heliospheric plasma}.  \jt{Astronomy \& Astrophysics}  \bvol{452}~(1),
  \pg{321--330}.

\bibitem[Bhattacharjee {\em et~al.\/}(2009)Bhattacharjee, Huang, Yang \&
  Rogers]{bhattacharjee2009fast}
{\sc \au{Bhattacharjee, A}, \au{Huang, Yi-Min}, \au{Yang, H} \& \au{Rogers, B}}
  \yr{2009}  \at{Fast reconnection in high-lundquist-number plasmas due to the
  plasmoid instability}.  \jt{Physics of Plasmas}  \bvol{16}~(11),
  \pg{112102}.

\bibitem[Bora {\em et~al.\/}(2021)Bora, Bhattacharyya \&
  Smolarkiewicz]{bora2021evolution}
{\sc \au{Bora, Kamlesh}, \au{Bhattacharyya, Ramit} \& \au{Smolarkiewicz,
  Piotr~K}} \yr{2021}  \at{Evolution of three-dimensional coherent structures
  in hall magnetohydrodynamics}.  \jt{The Astrophysical Journal}
  \bvol{906}~(2),  \pg{102}.

\bibitem[B{\"u}chner \& Elkina(2006)]{buchner2006anomalous}
{\sc \au{B{\"u}chner, J{\"o}rg} \& \au{Elkina, Nina}} \yr{2006}  \at{Anomalous
  resistivity of current-driven isothermal plasmas due to phase space
  structuring}.  \jt{Physics of plasmas}  \bvol{13}~(8),  \pg{082304}.

\bibitem[Burch {\em et~al.\/}(2016)Burch, Torbert, Phan, Chen, Moore, Ergun,
  Eastwood, Gershman, Cassak, Argall {\em et~al.\/}]{burch2016electron}
{\sc \au{Burch, JL}, \au{Torbert, RB}, \au{Phan, TD}, \au{Chen, L-J},
  \au{Moore, TE}, \au{Ergun, RE}, \au{Eastwood, JP}, \au{Gershman, DJ},
  \au{Cassak, PA}, \au{Argall, MR} \& \au{others}} \yr{2016}
  \at{Electron-scale measurements of magnetic reconnection in space}.
  \jt{Science}  \bvol{352}~(6290),  \pg{aaf2939}.

\bibitem[Cassak {\em et~al.\/}(2017)Cassak, Genestreti, Burch, Phan, Shay,
  Swisdak, Drake, Price, Eriksson, Ergun {\em et~al.\/}]{cassak2017effect}
{\sc \au{Cassak, PA}, \au{Genestreti, KJ}, \au{Burch, James~L}, \au{Phan, T-D},
  \au{Shay, MA}, \au{Swisdak, M}, \au{Drake, JF}, \au{Price, L}, \au{Eriksson,
  S}, \au{Ergun, RE} \& \au{others}} \yr{2017}  \at{The effect of a guide field
  on local energy conversion during asymmetric magnetic reconnection:
  Particle-in-cell simulations}.  \jt{Journal of Geophysical Research: Space
  Physics}  \bvol{122}~(11),  \pg{11--523}.

\bibitem[Chapman \& Ferraro(1940)]{chapman1940theory}
{\sc \au{Chapman, S} \& \au{Ferraro, VCA}} \yr{1940}  \at{The theory of the
  first phase of a geomagnetic storm}.  \jt{Terrestrial Magnetism and
  Atmospheric Electricity}  \bvol{45}~(3),  \pg{245--268}.

\bibitem[Chen {\em et~al.\/}(1997)Chen, Otto \& Lee]{chen1997tearing}
{\sc \au{Chen, Q}, \au{Otto, A} \& \au{Lee, LC}} \yr{1997}  \at{Tearing
  instability, kelvin-helmholtz instability, and magnetic reconnection}.
  \jt{Journal of Geophysical Research: Space Physics}  \bvol{102}~(A1),
  \pg{151--161}.

\bibitem[Chen \& Morrison(1990)]{chen1990resistive}
{\sc \au{Chen, XL} \& \au{Morrison, PJ}} \yr{1990}  \at{Resistive tearing
  instability with equilibrium shear flow}.  \jt{Physics of Fluids B: Plasma
  Physics}  \bvol{2}~(3),  \pg{495--507}.

\bibitem[Coppi {\em et~al.\/}(1966)Coppi, Greene \&
  Johnson]{coppi1966resistive}
{\sc \au{Coppi, Bruno}, \au{Greene, John~M} \& \au{Johnson, John~L}} \yr{1966}
  \at{Resistive instabilities in a diffuse linear pinch}.  \jt{Nuclear Fusion}
  \bvol{6}~(2),  \pg{101}.

\bibitem[Dahlburg {\em et~al.\/}(1997)Dahlburg, Boncinelli \&
  Einaudi]{dahlburg1997evolution}
{\sc \au{Dahlburg, RB}, \au{Boncinelli, P} \& \au{Einaudi, G}} \yr{1997}
  \at{The evolution of plane current--vortex sheets}.  \jt{Physics of Plasmas}
  \bvol{4}~(5),  \pg{1213--1226}.

\bibitem[Daughton {\em et~al.\/}(2006)Daughton, Scudder \&
  Karimabadi]{daughton2006fully}
{\sc \au{Daughton, William}, \au{Scudder, Jack} \& \au{Karimabadi, Homa}}
  \yr{2006}  \at{Fully kinetic simulations of undriven magnetic reconnection
  with open boundary conditions}.  \jt{Physics of Plasmas}  \bvol{13}~(7),
  \pg{072101}.

\bibitem[Dorfman {\em et~al.\/}(2013)Dorfman, Ji, Yamada, Yoo, Lawrence, Myers
  \& Tharp]{dorfman2013three}
{\sc \au{Dorfman, S}, \au{Ji, H}, \au{Yamada, M}, \au{Yoo, J}, \au{Lawrence,
  E}, \au{Myers, C} \& \au{Tharp, TD}} \yr{2013}  \at{Three-dimensional,
  impulsive magnetic reconnection in a laboratory plasma}.  \jt{Geophysical
  Research Letters}  \bvol{40}~(2),  \pg{233--238}.

\bibitem[Einaudi \& Rubini(1986)]{einaudi1986resistive}
{\sc \au{Einaudi, G} \& \au{Rubini, F}} \yr{1986}  \at{Resistive instabilities
  in a flowing plasma: I. inviscid case}.  \jt{The Physics of fluids}
  \bvol{29}~(8),  \pg{2563--2568}.

\bibitem[Faganello {\em et~al.\/}(2010)Faganello, Pegoraro, Califano \&
  Marradi]{faganello2010collisionless}
{\sc \au{Faganello, M}, \au{Pegoraro, Francesco}, \au{Califano, Francesco} \&
  \au{Marradi, L}} \yr{2010}  \at{Collisionless magnetic reconnection in the
  presence of a sheared velocity field}.  \jt{Physics of Plasmas}
  \bvol{17}~(6),  \pg{062102}.

\bibitem[Ferraro(1952)]{ferraro1952theory}
{\sc \au{Ferraro, VCA}} \yr{1952}  \at{On the theory of the first phase of a
  geomagnetic storm: A new illustrative calculation based on an idealised
  (plane not cylindrical) model field distribution}.  \jt{Journal of
  Geophysical Research}  \bvol{57}~(1),  \pg{15--49}.

\bibitem[Furth {\em et~al.\/}(1963)Furth, Killeen \&
  Rosenbluth]{furth1963finite}
{\sc \au{Furth, Harold~P}, \au{Killeen, John} \& \au{Rosenbluth, Marshall~N}}
  \yr{1963}  \at{Finite-resistivity instabilities of a sheet pinch}.  \jt{The
  physics of Fluids}  \bvol{6}~(4),  \pg{459--484}.

\bibitem[Gosling {\em et~al.\/}(1995)Gosling, Birn \& Hesse]{gosling1995three}
{\sc \au{Gosling, JT}, \au{Birn, J} \& \au{Hesse, M}} \yr{1995}
  \at{Three-dimensional magnetic reconnection and the magnetic topology of
  coronal mass ejection events}.  \jt{Geophysical research letters}
  \bvol{22}~(8),  \pg{869--872}.

\bibitem[Guo {\em et~al.\/}(2015)Guo, Liu, Daughton \& Li]{guo2015particle}
{\sc \au{Guo, Fan}, \au{Liu, Yi-Hsin}, \au{Daughton, William} \& \au{Li, Hui}}
  \yr{2015}  \at{Particle acceleration and plasma dynamics during magnetic
  reconnection in the magnetically dominated regime}.  \jt{The Astrophysical
  Journal}  \bvol{806}~(2),  \pg{167}.

\bibitem[Hofman(1975)]{hofman1975resistive}
{\sc \au{Hofman, I}} \yr{1975}  \at{Resistive tearing modes in a sheet pinch
  with shear flow}.  \jt{Plasma physics}  \bvol{17}~(2),  \pg{143}.

\bibitem[Hoshino \& Higashimori(2015)]{hoshino2015generation}
{\sc \au{Hoshino, Masahiro} \& \au{Higashimori, Katsuaki}} \yr{2015}
  \at{Generation of alfv{\'e}nic waves and turbulence in reconnection jets}.
  \jt{Journal of Geophysical Research: Space Physics}  \bvol{120}~(5),
  \pg{3715--3727}.

\bibitem[Huang {\em et~al.\/}(2016)Huang, Sahraoui, Retin{\`o}, Le~Contel,
  Yuan, Chasapis, Aunai, Breuillard, Deng, Zhou {\em et~al.\/}]{huang2016mms}
{\sc \au{Huang, SY}, \au{Sahraoui, Fouad}, \au{Retin{\`o}, Alessandro},
  \au{Le~Contel, Olivier}, \au{Yuan, ZG}, \au{Chasapis, Alexandros}, \au{Aunai,
  Nicolas}, \au{Breuillard, Hugo}, \au{Deng, XH}, \au{Zhou, Meng} \&
  \au{others}} \yr{2016}  \at{Mms observations of ion-scale magnetic island in
  the magnetosheath turbulent plasma}.  \jt{Geophysical Research Letters}
  \bvol{43}~(15),  \pg{7850--7858}.

\bibitem[Huang \& Bhattacharjee(2013)]{huang2013plasmoid}
{\sc \au{Huang, Yi-Min} \& \au{Bhattacharjee, A}} \yr{2013}  \at{Plasmoid
  instability in high-lundquist-number magnetic reconnection}.  \jt{Physics of
  Plasmas}  \bvol{20}~(5),  \pg{055702}.

\bibitem[Ji \& Daughton(2011)]{ji2011phase}
{\sc \au{Ji, Hantao} \& \au{Daughton, William}} \yr{2011}  \at{Phase diagram
  for magnetic reconnection in heliophysical, astrophysical, and laboratory
  plasmas}.  \jt{Physics of Plasmas}  \bvol{18}~(11),  \pg{111207}.

\bibitem[Landi {\em et~al.\/}(2015)Landi, Del~Zanna, Papini, Pucci \&
  Velli]{landi2015resistive}
{\sc \au{Landi, Simone}, \au{Del~Zanna, Luca}, \au{Papini, Emanuele},
  \au{Pucci, Fulvia} \& \au{Velli, Marco}} \yr{2015}  \at{Resistive
  magnetohydrodynamics simulations of the ideal tearing mode}.  \jt{The
  Astrophysical Journal}  \bvol{806}~(1),  \pg{131}.

\bibitem[Lane {\em et~al.\/}(2021)Lane, Grocott, Case \&
  Walach]{lane2021dynamics}
{\sc \au{Lane, James~H}, \au{Grocott, Adrian}, \au{Case, Nathan~A} \&
  \au{Walach, Maria-Theresia}} \yr{2021} Dynamics of variable dusk--dawn flow
  associated with magnetotail current sheet flapping.  \bt{In {\em Annales
  Geophysicae\/}}, ,  \vol{vol.~39},  \pg{pp. 1037--1053}. Copernicus GmbH.

\bibitem[Lee {\em et~al.\/}(1988)Lee, Wang, Wei \& Tsurutani]{lee1988streaming}
{\sc \au{Lee, LC}, \au{Wang, S}, \au{Wei, CQ} \& \au{Tsurutani, BT}} \yr{1988}
  \at{Streaming sausage, kink and tearing instabilities in a current sheet with
  applications to the earth's magnetotail}.  \jt{Journal of Geophysical
  Research: Space Physics}  \bvol{93}~(A7),  \pg{7354--7365}.

\bibitem[Loureiro {\em et~al.\/}(2007)Loureiro, Schekochihin \&
  Cowley]{loureiro2007instability}
{\sc \au{Loureiro, NF}, \au{Schekochihin, AA} \& \au{Cowley, SC}} \yr{2007}
  \at{Instability of current sheets and formation of plasmoid chains}.
  \jt{Physics of Plasmas}  \bvol{14}~(10),  \pg{100703}.

\bibitem[Lu {\em et~al.\/}(2019)Lu, Angelopoulos, Artemyev, Pritchett, Liu,
  Runov, Tenerani, Shi \& Velli]{lu2019turbulence}
{\sc \au{Lu, San}, \au{Angelopoulos, V}, \au{Artemyev, AV}, \au{Pritchett, PL},
  \au{Liu, J}, \au{Runov, A}, \au{Tenerani, A}, \au{Shi, C} \& \au{Velli, M}}
  \yr{2019}  \at{Turbulence and particle acceleration in collisionless magnetic
  reconnection: Effects of temperature inhomogeneity across pre-reconnection
  current sheet}.  \jt{The Astrophysical Journal}  \bvol{878}~(2),  \pg{109}.

\bibitem[Ma {\em et~al.\/}(2018)Ma, Chen, Zhang \& Yu]{ma2018effective}
{\sc \au{Ma, Zhi-Wei}, \au{Chen, Tong}, \au{Zhang, HW} \& \au{Yu, MY}}
  \yr{2018}  \at{Effective resistivity in collisionless magnetic reconnection}.
   \jt{Scientific Reports}  \bvol{8}~(1),  \pg{1--6}.

\bibitem[Masuda {\em et~al.\/}(1994)Masuda, Kosugi, Hara, Tsuneta \&
  Ogawara]{masuda1994loop}
{\sc \au{Masuda, S}, \au{Kosugi, T}, \au{Hara, H}, \au{Tsuneta, S} \&
  \au{Ogawara, Y}} \yr{1994}  \at{A loop-top hard x-ray source in a compact
  solar flare as evidence for magnetic reconnection}.  \jt{Nature}
  \bvol{371}~(6497),  \pg{495--497}.

\bibitem[Ofman {\em et~al.\/}(1991)Ofman, Chen, Morrison \&
  Steinolfson]{ofman1991resistive}
{\sc \au{Ofman, L}, \au{Chen, XL}, \au{Morrison, PJ} \& \au{Steinolfson, RS}}
  \yr{1991}  \at{Resistive tearing mode instability with shear flow and
  viscosity}.  \jt{Physics of Fluids B: Plasma Physics}  \bvol{3}~(6),
  \pg{1364--1373}.

\bibitem[Osman {\em et~al.\/}(2014)Osman, Matthaeus, Gosling, Greco, Servidio,
  Hnat, Chapman \& Phan]{osman2014magnetic}
{\sc \au{Osman, KT}, \au{Matthaeus, WH}, \au{Gosling, JT}, \au{Greco, A},
  \au{Servidio, S}, \au{Hnat, B}, \au{Chapman, Sandra~C} \& \au{Phan, TD}}
  \yr{2014}  \at{Magnetic reconnection and intermittent turbulence in the solar
  wind}.  \jt{Physical Review Letters}  \bvol{112}~(21),  \pg{215002}.

\bibitem[Papini {\em et~al.\/}(2019)Papini, Landi \& Del~Zanna]{papini2019fast}
{\sc \au{Papini, Emanuele}, \au{Landi, Simone} \& \au{Del~Zanna, Luca}}
  \yr{2019}  \at{Fast magnetic reconnection: secondary tearing instability and
  role of the hall term}.  \jt{The Astrophysical Journal}  \bvol{885}~(1),
  \pg{56}.

\bibitem[Paris \& Sy(1983)]{paris1983influence}
{\sc \au{Paris, RB} \& \au{Sy, WN-C}} \yr{1983}  \at{Influence of equilibrium
  shear flow along the magnetic field on the resistive tearing instability}.
  \jt{The Physics of fluids}  \bvol{26}~(10),  \pg{2966--2975}.

\bibitem[Paris {\em et~al.\/}(1993)Paris, Wood \& Stewart]{paris1993effects}
{\sc \au{Paris, RB}, \au{Wood, AD} \& \au{Stewart, S}} \yr{1993}  \at{Effects
  of equilibrium flow on the resistive tearing mode}.  \jt{Physics of Fluids B:
  Plasma Physics}  \bvol{5}~(3),  \pg{1027--1029}.

\bibitem[Parker(1957)]{parker1957sweet}
{\sc \au{Parker, Eugene~N}} \yr{1957}  \at{Sweet's mechanism for merging
  magnetic fields in conducting fluids}.  \jt{Journal of Geophysical Research}
  \bvol{62}~(4),  \pg{509--520}.

\bibitem[Pucci {\em et~al.\/}(2020)Pucci, Singh, Tenerani \&
  Velli]{pucci2020tearing}
{\sc \au{Pucci, Fulvia}, \au{Singh, K Alkendra~P}, \au{Tenerani, Anna} \&
  \au{Velli, Marco}} \yr{2020}  \at{Tearing modes in partially ionized
  astrophysical plasma}.  \jt{The Astrophysical Journal Letters}
  \bvol{903}~(1),  \pg{L19}.

\bibitem[Pucci \& Velli(2013)]{pucci2013reconnection}
{\sc \au{Pucci, Fulvia} \& \au{Velli, Marco}} \yr{2013}  \at{Reconnection of
  quasi-singular current sheets: the “ideal” tearing mode}.  \jt{The
  Astrophysical Journal Letters}  \bvol{780}~(2),  \pg{L19}.

\bibitem[Pucci {\em et~al.\/}(2017)Pucci, Velli \& Tenerani]{pucci2017fast}
{\sc \au{Pucci, Fulvia}, \au{Velli, Marco} \& \au{Tenerani, Anna}} \yr{2017}
  \at{Fast magnetic reconnection:“ideal” tearing and the hall effect}.
  \jt{The Astrophysical Journal}  \bvol{845}~(1),  \pg{25}.

\bibitem[Pucci {\em et~al.\/}(2018)Pucci, Velli, Tenerani \&
  Del~Sarto]{pucci2018onset}
{\sc \au{Pucci, Fulvia}, \au{Velli, Marco}, \au{Tenerani, Anna} \&
  \au{Del~Sarto, Daniele}} \yr{2018}  \at{Onset of fast “ideal” tearing in
  thin current sheets: dependence on the equilibrium current profile}.
  \jt{Physics of Plasmas}  \bvol{25}~(3),  \pg{032113}.

\bibitem[R{\'e}ville {\em et~al.\/}(2022)R{\'e}ville, Fargette, Rouillard,
  Lavraud, Velli, Strugarek, Parenti, Brun, Shi, Kouloumvakos {\em
  et~al.\/}]{reville2022flux}
{\sc \au{R{\'e}ville, V}, \au{Fargette, N}, \au{Rouillard, AP}, \au{Lavraud,
  B}, \au{Velli, M}, \au{Strugarek, A}, \au{Parenti, S}, \au{Brun, AS},
  \au{Shi, C}, \au{Kouloumvakos, A} \& \au{others}} \yr{2022}  \at{Flux rope
  and dynamics of the heliospheric current sheet-study of the parker solar
  probe and solar orbiter conjunction of june 2020}.  \jt{Astronomy \&
  Astrophysics}  \bvol{659},  \pg{A110}.

\bibitem[R{\'e}ville {\em et~al.\/}(2020)R{\'e}ville, Velli, Rouillard,
  Lavraud, Tenerani, Shi \& Strugarek]{reville2020tearing}
{\sc \au{R{\'e}ville, Victor}, \au{Velli, Marco}, \au{Rouillard, Alexis~P},
  \au{Lavraud, Benoit}, \au{Tenerani, Anna}, \au{Shi, Chen} \& \au{Strugarek,
  Antoine}} \yr{2020}  \at{Tearing instability and periodic density
  perturbations in the slow solar wind}.  \jt{The Astrophysical Journal
  Letters}  \bvol{895}~(1),  \pg{L20}.

\bibitem[Shi {\em et~al.\/}(2021)Shi, Artemyev, Velli \&
  Tenerani]{shi2021stability}
{\sc \au{Shi, Chen}, \au{Artemyev, Anton}, \au{Velli, Marco} \& \au{Tenerani,
  Anna}} \yr{2021}  \at{Stability of the magnetotail current sheet with normal
  magnetic field and field-aligned plasma flows}.  \jt{Journal of Geophysical
  Research: Space Physics}  \bvol{126}~(11),  \pg{e2021JA029711}.

\bibitem[Shi {\em et~al.\/}(2019)Shi, Tenerani, Velli \& Lu]{shi2019fast}
{\sc \au{Shi, Chen}, \au{Tenerani, Anna}, \au{Velli, Marco} \& \au{Lu, San}}
  \yr{2019}  \at{Fast recursive reconnection and the hall effect: Hall-mhd
  simulations}.  \jt{The Astrophysical Journal}  \bvol{883}~(2),  \pg{172}.

\bibitem[Shi {\em et~al.\/}(2020)Shi, Velli, Pucci, Tenerani \&
  Innocenti]{shi2020oblique}
{\sc \au{Shi, Chen}, \au{Velli, Marco}, \au{Pucci, Fulvia}, \au{Tenerani, Anna}
  \& \au{Innocenti, Maria~Elena}} \yr{2020}  \at{Oblique tearing mode
  instability: guide field and hall effect}.  \jt{The Astrophysical Journal}
  \bvol{902}~(2),  \pg{142}.

\bibitem[Shi {\em et~al.\/}(2018)Shi, Velli \& Tenerani]{shi2018marginal}
{\sc \au{Shi, Chen}, \au{Velli, Marco} \& \au{Tenerani, Anna}} \yr{2018}
  \at{Marginal stability of sweet--parker type current sheets at low lundquist
  numbers}.  \jt{The Astrophysical Journal}  \bvol{859}~(2),  \pg{83}.

\bibitem[Shibata \& Tanuma(2001)]{shibata2001plasmoid}
{\sc \au{Shibata, Kazunari} \& \au{Tanuma, Syuniti}} \yr{2001}
  \at{Plasmoid-induced-reconnection and fractal reconnection}.  \jt{Earth,
  Planets and Space}  \bvol{53}~(6),  \pg{473--482}.

\bibitem[Sweet(1958)]{sweet1958electromagnetic}
{\sc \au{Sweet, PA}} \yr{1958} Electromagnetic phenomena in cosmical physics.
  \bt{In {\em IAU Symp. 6\/}}, ,  \vol{vol. 123}. Kluwer Academic Publishers.

\bibitem[Tajima \& Shibata(2018)]{tajima2018plasma}
{\sc \au{Tajima, Toshiki} \& \au{Shibata, Kazunari}} \yr{2018} {\em Plasma
  astrophysics\/}.  \publ{CRC Press}.

\bibitem[Tassi {\em et~al.\/}(2014)Tassi, Grasso \& Comisso]{tassi2014linear}
{\sc \au{Tassi, Emanuele}, \au{Grasso, Daniela} \& \au{Comisso, Luca}}
  \yr{2014}  \at{Linear stability analysis of collisionless reconnection in the
  presence of an equilibrium flow aligned with the guide field}.  \jt{The
  European Physical Journal D}  \bvol{68}~(4),  \pg{1--11}.

\bibitem[Tenerani {\em et~al.\/}(2015{\natexlab{{\em a\/}}})Tenerani, Rappazzo,
  Velli \& Pucci]{tenerani2015tearing}
{\sc \au{Tenerani, Anna}, \au{Rappazzo, Antonio~Franco}, \au{Velli, Marco} \&
  \au{Pucci, Fulvia}} \yr{2015{\natexlab{{\em a\/}}}}  \at{The tearing mode
  instability of thin current sheets: the transition to fast reconnection in
  the presence of viscosity}.  \jt{The Astrophysical Journal}  \bvol{801}~(2),
  \pg{145}.

\bibitem[Tenerani {\em et~al.\/}(2015{\natexlab{{\em b\/}}})Tenerani, Velli,
  Rappazzo \& Pucci]{tenerani2015magnetic}
{\sc \au{Tenerani, Anna}, \au{Velli, Marco}, \au{Rappazzo, Antonio~Franco} \&
  \au{Pucci, Fulvia}} \yr{2015{\natexlab{{\em b\/}}}}  \at{Magnetic
  reconnection: Recursive current sheet collapse triggered by “ideal”
  tearing}.  \jt{The Astrophysical Journal Letters}  \bvol{813}~(2),  \pg{L32}.

\bibitem[Torbert {\em et~al.\/}(2018)Torbert, Burch, Phan, Hesse, Argall,
  Shuster, Ergun, Alm, Nakamura, Genestreti {\em
  et~al.\/}]{torbert2018electron}
{\sc \au{Torbert, RB}, \au{Burch, JL}, \au{Phan, TD}, \au{Hesse, M},
  \au{Argall, MR}, \au{Shuster, J}, \au{Ergun, RE}, \au{Alm, Love},
  \au{Nakamura, R}, \au{Genestreti, KJ} \& \au{others}} \yr{2018}
  \at{Electron-scale dynamics of the diffusion region during symmetric magnetic
  reconnection in space}.  \jt{Science}  \bvol{362}~(6421),  \pg{1391--1395}.

\bibitem[Virtanen {\em et~al.\/}(2020)Virtanen, Gommers, Oliphant, Haberland,
  Reddy, Cournapeau, Burovski, Peterson, Weckesser, Bright {\em
  et~al.\/}]{virtanen2020scipy}
{\sc \au{Virtanen, Pauli}, \au{Gommers, Ralf}, \au{Oliphant, Travis~E},
  \au{Haberland, Matt}, \au{Reddy, Tyler}, \au{Cournapeau, David},
  \au{Burovski, Evgeni}, \au{Peterson, Pearu}, \au{Weckesser, Warren},
  \au{Bright, Jonathan} \& \au{others}} \yr{2020}  \at{Scipy 1.0: fundamental
  algorithms for scientific computing in python}.  \jt{Nature methods}
  \bvol{17}~(3),  \pg{261--272}.

\bibitem[Wallace {\em et~al.\/}(2010)Wallace, Harra, van Driel-Gesztelyi, Green
  \& Matthews]{wallace2010pre}
{\sc \au{Wallace, AJ}, \au{Harra, LK}, \au{van Driel-Gesztelyi, L}, \au{Green,
  LM} \& \au{Matthews, SA}} \yr{2010}  \at{Pre-flare flows in the corona}.
  \jt{Solar Physics}  \bvol{267}~(2),  \pg{361--375}.

\bibitem[Wang {\em et~al.\/}(1988{\natexlab{{\em a\/}}})Wang, Lee \&
  Wei]{wang1988streaming}
{\sc \au{Wang, S}, \au{Lee, LC} \& \au{Wei, CQ}} \yr{1988{\natexlab{{\em
  a\/}}}}  \at{Streaming tearing instability in the current sheet with a
  super-alfvenic flow}.  \jt{The Physics of fluids}  \bvol{31}~(6),
  \pg{1544--1548}.

\bibitem[Wang {\em et~al.\/}(1988{\natexlab{{\em b\/}}})Wang, Lee, Wei \&
  Akasofu]{wang1988mechanism}
{\sc \au{Wang, S}, \au{Lee, LC}, \au{Wei, CQ} \& \au{Akasofu, S-I}}
  \yr{1988{\natexlab{{\em b\/}}}}  \at{A mechanism for the formation of
  plasmoids and kink waves in the heliospheric current sheet}.  \jt{Solar
  physics}  \bvol{117}~(1),  \pg{157--169}.

\bibitem[Yamada {\em et~al.\/}(1994)Yamada, Levinton, Pomphrey, Budny, Manickam
  \& Nagayama]{yamada1994investigation}
{\sc \au{Yamada, M}, \au{Levinton, FM}, \au{Pomphrey, N}, \au{Budny, R},
  \au{Manickam, J} \& \au{Nagayama, Y}} \yr{1994}  \at{Investigation of
  magnetic reconnection during a sawtooth crash in a high-temperature tokamak
  plasma}.  \jt{Physics of plasmas}  \bvol{1}~(10),  \pg{3269--3276}.

\end{thebibliography}

\end{document}